\documentclass[twocolumn,preprintnumbers,showpacs,pre,floatfix,amsmath,amssymb,notitlepage,superscriptaddress]{revtex4-1}
\usepackage{epsfig}
\usepackage{subfigure}
\usepackage{amsmath}
\usepackage{color}
\usepackage{amssymb}
\usepackage{setspace}
\usepackage{graphicx}
\usepackage{dcolumn}
\usepackage{bm}
\usepackage{times}
\usepackage{enumerate}
\usepackage{footnote}
\input epsf

\newcommand{\drt}[1]{{\color{black}#1}}

\graphicspath{{figures/}}

\begin{document}

\author{Dane Taylor}
\email{dane.r.taylor@gmail.com}\affiliation{Carolina Center for Interdisciplinary Applied Mathematics, Department of Mathematics, University of North Carolina, Chapel Hill, NC 27599, USA}\affiliation{\drt{Department of Mathematics, University at Buffalo,  State University of New York (SUNY), Buffalo, NY 14260, USA}}
\author{Rajmonda S. Caceres}\affiliation{Lincoln Laboratory, Massachusetts Institute of Technology, Lexington, MA 02420, USA}
\author{Peter J. Mucha}\affiliation{Carolina Center for Interdisciplinary Applied Mathematics, Department of Mathematics, University of North Carolina, Chapel Hill, NC 27599, USA}

\title{
\drt{Super-resolution community detection for layer-aggregated multilayer networks}
}

\begin{abstract}
\drt{
Applied network science often involves preprocessing network data before applying a network-analysis method, and  there is typically a theoretical disconnect between these steps.
For example, it is common  to aggregate time-varying network data into windows prior to analysis, and the tradeoffs of this preprocessing are not well understood.
Focusing on the problem of detecting small communities in multilayer networks, we study the effects of layer aggregation by developing random-matrix theory for modularity matrices associated with layer-aggregated networks with
%
$N$ nodes and $L$ layers, which are drawn from an ensemble of Erd\H{o}s--R\'enyi networks. 
We study phase transitions in which eigenvectors localize onto communities (allowing their detection) and which occur for a given community provided its size surpasses a {detectability limit} $K^*$.
When layers are aggregated via a summation, we obtain $K^*\varpropto \mathcal{O}(\sqrt{NL}/T)$, where $T$ is the number of layers across which the community persists. 
Interstingly, if $T$ is allowed to vary with $L$ then summation-based layer aggregation enhances small-community detection even if the community persists across a vanishing fraction of layers, provided that $T/L$ decays  more slowly than $ \mathcal{O}(L^{-1/2})$.
%
%
%
%
Moreover, we find that thresholding the summation can in some cases cause $K^*$ to decay exponentially, decreasing by  orders of magnitude in 
a phenomenon  we call 
\emph{super-resolution community detection}.
That is, layer aggregation with thresholding is a \emph{nonlinear data filter} enabling detection of communities that are otherwise too small to detect. 
Importantly,  different thresholds generally enhance the detectability of communities having different properties, illustrating that community detection can be obscured if one analyzes network data using a single threshold.
}

\end{abstract}

\pacs{89.75.Fb, 02.70.Hm, 64.60.aq}

\maketitle

\section{Introduction}

Network-based modeling provides a powerful framework for analyzing high-dimensional data sets and complex systems~\cite{Newman2003}. Often, a network is best represented by a set of network layers that encode different types of interactions, such as categorical social ties \cite{Lewis} or a network at different instances in time~\cite{Holme2012}, and an important pursuit involves extending network theory to the multilayer setting \cite{Kivela2014,Boccaletti2014}. Sometimes, however, a multilayer framework can require too much computational overhead or can represent an over-modeling (e.g., when the layers are correlated, either in terms of the edge overlap \cite{Menichetti2014} or other properties \cite{Domenico2015,Stanley2015,Kleineberg2016}), and it can be beneficial to aggregate layers~\cite{Stanley2015,Taylor2016,Nayar2015}. In particular, aggregation provides a crucial step for analyzing temporal network data, which is often binned into time windows \cite{Mucha2010,Bassett2011}. 
\drt{Layer aggregation and other types of network preprocessing (e.g., sparsification \cite{chung}, network inference \cite{Hill2016} and de-noising \cite{Clauset2008,Newman2017}) can greatly influence the resulting network structure, which in turn influences outcomes of network analyses and their many applications.
In general, there remains a significant need for improved theoretical understanding for how such network preprocessing influences network-analysis methodology.
}


\begin{figure}[t]
\centering
\epsfig{file = 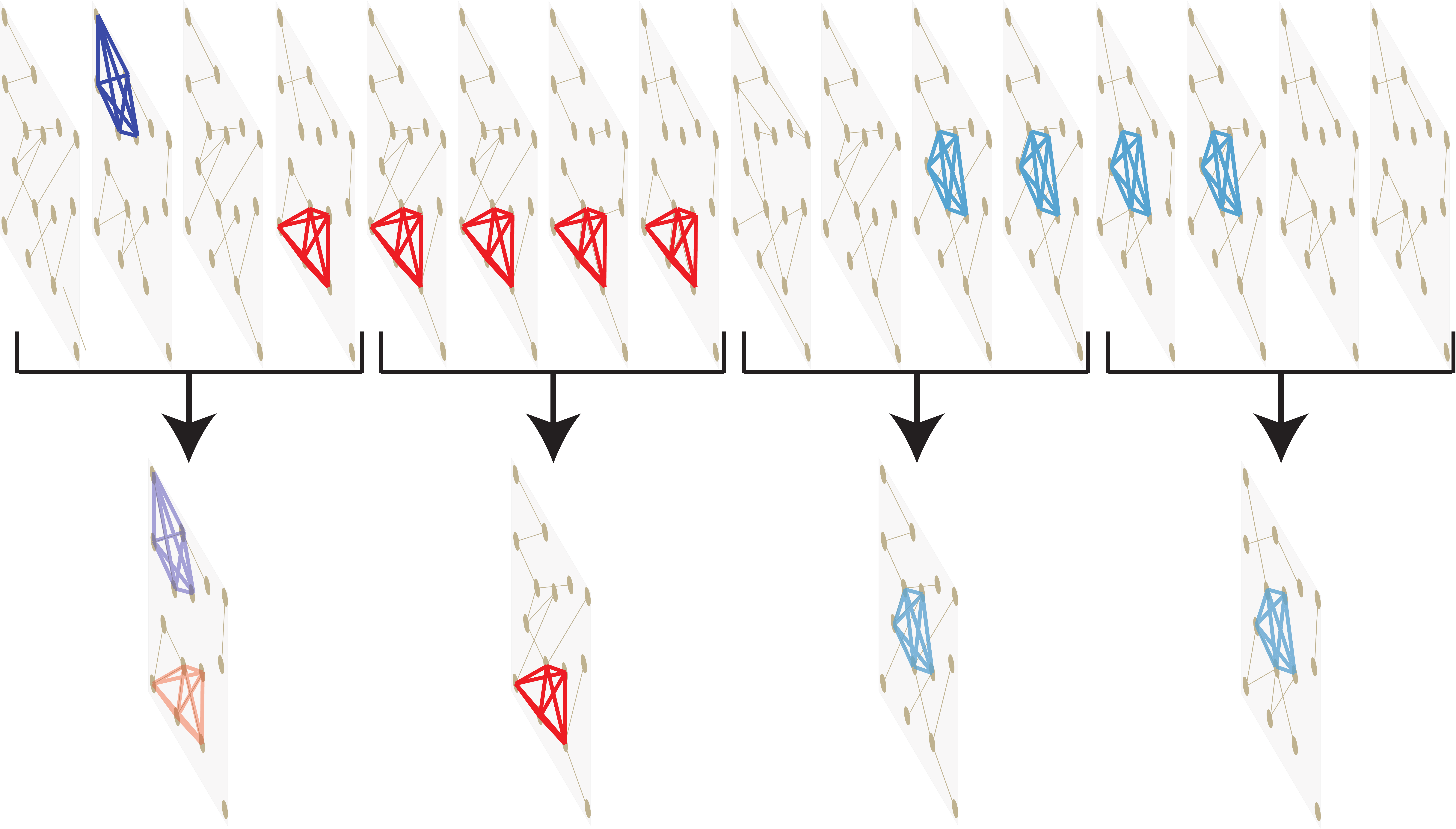, clip =,width=1\linewidth }
\caption{
\drt{
Preprocessing  networks (including multilayer representations of temporal networks) often involves aggregating 
network data into bins (or time windows).
We study how many layers must contain a community in order for aggregation to enhance its detection  and introduce \emph{layer aggregation with thresholding}  as a nonlinear data filter enabling super-resolution community detection.
}
} 
\label{fig:1_windows}
\end{figure}

\drt{We study the effects of layer aggregation on  community detection,  one of the  widely used methods for studying social, biological and physical networks \cite{Sales2007,Rosvall2008,Lancichinetti2009,Fortunato}.
%
Communities are typically studied as dense subgraphs and can} represent, for example, coordinating neurons in the brain \cite{Bassett2011} or a social clique \cite{moody} in a social network. 
(Hereafter, we will  restrict our usage of the term `clique' to the graph-theoretical meaning of a subgraph with all-to-all coupling.) Of particular interest is the detection of small-scale communities, 
which is a paradigmatic pursuit for anomaly detection 
within the fields of 
signal processing and cybersecurity \cite{Alon1998,Nadakuditi2012,Miller2015,Mavroeidis,Ding2012,Chen2013}. 
\drt{In this context, small communities can represent anomalous events 
such as attacks  \cite{Mavroeidis}, intrusions \cite{Ding2012}, and fraud \cite{Chen2013}.}

\drt{Given these and many other  applications, there is great interest to understand fundamental limitations on community detection 
\cite{Alon1998,Nadakuditi2012,Miller2015,Nayar2015,Decelle_2011,Nadakuditi_2012,Radicchi_2013,Peixoto2013,Chen2015,Ghasemian2015,Fortunato2007,Kawamoto_2015}. We highlight recent detectability results for multilayer \cite{Nayar2015,Taylor2016} and temporal networks \cite{Ghasemian2015}. It's worth noting that much of the detectability research has focused on} large-scale communities whose sizes are $O(N)$, where $N$ is the number of nodes in the network \cite{Decelle_2011,Nadakuditi_2012,Radicchi_2013,Peixoto2013,Chen2015,Ghasemian2015,Kawamoto_2015}, and the phase transitions are typically driven by varying the prevalence (e.g., edge density) of the communities. In contrast, detectability phase transitions for small communities can \drt{also} be onset by varying their size $K$ \cite{Alon1998,Nadakuditi2012,Miller2015,Nayar2015} \drt{and are thus a type of resolution limit \cite{Fortunato2007}.
For example, a planted clique  in a single-layer Erd\H{o}s-R\'enyi network is detectable via a spectral analysis only if its size $K$ surpasses a detectability limit $K^*\varpropto O(\sqrt{N})$~\cite{Alon1998}, in which case a dominant eigenvector localizes onto the clique.}
\drt{
Extending previous research for  the detectability of a  clique planted in single-layer networks \cite{Alon1998,Nadakuditi2012,Miller2015} and a clique that persists across all layers of a multilayer network \cite{Nayar2015}, 
herein we study the detectability of small communities (including, but not limit to, cliques) planted in a subset of layers in a multilayer network.}
%
%

\drt{With the application of detecting small communities in mind, we study the effects of layer aggregation as a network preprocessing step. We first ask a foundational question: Across how many layers must a community persist in order for layer-aggregation  to  benefit  detection.}
\drt{To this end, we study a multilayer network model in which small communities are} 
hidden in network layers generated as Erd\H{o}s--R\'enyi (ER) networks with $N$ nodes and $L$ \drt{layers} with (possibly)  heterogeneous edge probabilities.
\drt{We  study  detectability phase transitions wherein  eigenvectors localize onto communities, which we analyze by developing} 
random matrix theory for the eigenvectors of modularity matrices associated with an aggregation of the layers. 
When the aggregation is given by summation of the adjacency matrices, the detectability phase transition occurs when a community's size $K\ll N$ surpasses a critical value $K^*\varpropto \sqrt{NL}/T$, where
 \drt{$T$ is the number of layers across which a community persists.
Note that if $T$ depends on $L$ then summation-based layer aggregation benefits small-community detection even if the fraction $T/L$ of layers containing the community vanishes, provided that the fraction decays more slowly than $\mathcal{O}(L^{-1/2})$.

We additionally study network preprocessing via thresholding---that is, we threshold a summation of layers' adjacency matrices at some value $\tilde{L}$ so that there exists an unweighted edge between two nodes in the aggregated network if and only if there exists at least $\tilde{L}$ edges between them across the $L$ layers.
While it is well known that thresholding can be used to simultaneously sparsify and dichotomize a network, here we introduce thresholding as a \emph{nonlinear data filter} for enhancing small-community detection.
Specifically, we find that thresholding can  in some cases reduce $ {K}^*$ by orders of magnitude, revealing communities that are otherwise too small to detect.
%
We call this phenomenon \emph{super-resolution community detection} and show for clique detection in sparse networks that $ {K}^*$ decays exponentially with $\sqrt{L}/T$ for threshold $\tilde{L}=T$.
Importantly, we find that  different thresholds enhance the detection of communities with different properties (e.g., size and edge density), illustrating how community structure can be obscured if one uses a single threshold, which is an important insight for network preprocessing in general. 
}



\drt{
The remainder of this paper is organized as follows. 
In Sec.~\ref{sec:model}, we further specify our model.
In Sec.~\ref{sec:analysis}, we study the effects of layer aggregation on detectability phase transitions characterized by  eigenvector localization.
In Sec.~\ref{sec:time}, we highlight  implications of our  findings with a numerical experiment involving small-community detection in a temporal network.
We provide a discussion in Sec.~\ref{sec:discussion}}

\section{Model}\label{sec:model}

\subsection{Multilayer Networks with Planted Small Communities}

We generate $L$ network layers with $N$ nodes so that each layer $l\in\{1,\dots,L\}$ is an ER random graph with edge probability $p_l\in(0,1)$, which is allowed to vary across the layers. 
\drt{
We plant $R$ communities via the following process.
For $r\in\{1\dots,R\}$,  uniformly at random we select a set $\mathcal{T}_r\subset\{1,\dots,L\}$ of layers and a set $\mathcal{K}_r\subset \mathcal{V}=\{1,\dots,N\}$ of nodes, and we define an edge probability $\rho_r$. Variable $K_r=|\mathcal{K}_r|\ll N$ denotes the size of community $r$ and we refer to $T_r=|\mathcal{T}_r|$ as its \emph{persistence} across network layers.
Then for each $r$, we construct a dense subgraph between nodes $\mathcal{K}_r$ in layers $\mathcal{T}_r$ by first removing edges between them occurring under the ER model and creating new edges with probability $\rho_r$. 
To ensure that the communities are denser than the remaining network, we assume $\rho_r>\langle p_l\rangle$, where $\langle \cdot\rangle$} denotes the mean value across all layers. We allow self edges in both the ER model and the \drt{planted communities.} We note that the layers are not required to have a particular ordering, and the community is not restricted only to consecutive layers.  
%
Moreover, we restrict our study to non-overlapping communities by assuming that the communities involve different nodes so that $\mathcal{K}_r\cap \mathcal{K}_s=0$ for any $r\not=s$. 
We leave open the study of eigenvector localization in the case of overlapping communities.
\drt{Finally, we assume $\sum_r K_r \ll N$ so that only a small fraction of nodes are involved in communities, making them anomalous structures.}


\subsection{Layer-Aggregation Methods}

\drt{We find that layer aggregation is a  preprocessing step for multilayer  networks that can be used to reduce data size and/or as a  data filter to benefit  network-analysis outcomes such as community detection.}
Following the approach in \cite{Taylor2016}, we study two methods for aggregating layers of a multilayer network:
\begin{enumerate}
\item[(i).] 
The \emph{summation} network corresponds to the weighted adjacency matrix ${\bf \overline{A}} = \sum_l {\bf A}^{(l)}$, where ${\bf A}^{(l)}$ denotes the symmetric adjacency matrix encoding each \drt{network} layer $l\in\{1,\dots,L\}$. 
\item[(ii).] 
The family of \emph{thresholded} networks represented by unweighted adjacency matrices $\{{\bf\hat{A}}^{(\tilde{L})}\}$ are obtained by applying a threshold $\tilde{L}\in\{1,\dots,L\}$ to the entries $\{\overline{A}_{ij}\}$ of matrix ${\bf \overline{A}}$,
\begin{align}
\hat{A}^{(\tilde{L})}_{ij} = \left\{ \begin{array}{rl}1,&\text{if }  \overline{A}_{ij}\ge \tilde{L}\\0,&\text{otherwise.}\end{array}\right.
\end{align}
\end{enumerate}
\drt{Note that thresholding dichotomizes the network, and one can vary $\tilde{L}$ to tunably sparsify the network.}

\section{Detectability of Small Communities with Eigenvector Localization}\label{sec:analysis}

\drt{We now develop random matrix theory to analyze how layer aggregation  affects small-community detection.
In Sec.~\ref{sec:sum}, we present results for aggregation by summation, studying the fraction of layers that must contain a community in order for  layer aggregation to enhance  detection.
In Sec.~\ref{sec:thresh}, we present results for layer aggregation with thresholding, highlighting that certain threshold values can  yield super-resolution community detection.}

\subsection{Layer Aggregation via Summation}\label{sec:sum}

\subsubsection{Random Matrix Theory for Modularity Matrices}

We first describe the statistical properties of matrix entries $\{\overline{A}_{ij}\}$. 
\drt{For  edges $(i,j)\not \in \cup_r \{\mathcal{K}_r\times \mathcal{K}_r\}$,}
$\{\overline{A}_{ij}\}$ are independent and identically distributed (i.i.d.) random variables following a Poisson binomial distribution, $P(\overline{A}_{ij}=a) = f_{PB}(a;L,\{p_l\})$, where
\begin{equation}
 f_{PB}(a;L,\{p_l\}) = \sum_{\mathcal{S}\in\mathcal{S}_a} \prod_{l\in\mathcal{S}} {p_l} \prod_{m\in \{1,\dots,L\}\setminus\mathcal{S}} (1-p_m) ,\label{eq:Entry_dist}
\end{equation}
and $\mathcal{S}_a$ denotes the set of $\left(L\atop a\right)$ different subsets of layers $\{1,\dots,L\}$ that have cardinality $a$ (i.e., $\mathcal{S}_1=\{\{1\},\{2\},\dots\}$, $\mathcal{S}_2=\{\{1,2\},\{1,3\},\dots\}$, and so on).
We note that $ f_{PB}(a;L,\{p_l\})$ has mean $L\langle p_l\rangle$ and variance $L\langle  p_l(1-p_l)\rangle$. 
When the edge probability is identical across the layers (i.e., $p_l=p$), then Eq.~\eqref{eq:Entry_dist} simplifies to the binomial distribution, 
\begin{equation}
f(a;L,p)  =  \left(\footnotesize{\begin{array}{c}L\\ a\end{array}}\right) {p^a} (1-p)^{L-a}, \label{eq:Entry_dist2}
\end{equation}
with mean $ L p$ and variance $L p(1-p)$. 

\drt{For within-community edges $(i,j)\in\{\mathcal{K}_r\times\mathcal{K}_r\}$ associated with community $r$, the entries $\{\overline{A}_{ij}\}$ are i.i.d.~random variables following $ f_{PB}(a;L,\{q^{(r)}_l\})$, where $q^{(r)}_l=\rho_r$ for $l\in\mathcal{T}_r$ and otherwise $q^{(r)}_l=p_l$. It follows that the entries have mean $T_r\rho_r + \sum_{l\in\{1,\dots,L\}\setminus \mathcal{T}_r} p_l $ and variance $T_r\rho_r(1-\rho_r) + \sum_{l\in\{1,\dots,L\}\setminus \mathcal{T}_r} p_l(1-p_l)$. Because the layers $\mathcal{T}_r$ are selected uniformly at random, the expected mean and variance across all possible choices for $\mathcal{T}_r$ are given by $T_r\rho_r + (L-T_r)\langle p_l\rangle$ and $T_r\rho_r(1-\rho_r)+(L-T_r)\langle p_l(1-p_l)\rangle$, respectively.}

We now study the spectra of the modularity matrix \cite{Newman_2004}
\begin{align}
{\bf\overline{B}}={\bf\overline{A}} - L\langle p_i \rangle {\bf1}{\bf1}^T, \label{eq:mod}
\end{align}  
based on an ER null model in which each edge has expected weight $L\langle p_i\rangle$. 
Importantly, this null model does not use knowledge that edges $(i,j)$ between nodes \drt{$i,j\in\mathcal{K}_r$} have different expected edge probability \drt{(i.e., $T_r\rho + (L-T_r)\langle p_i\rangle$ versus $L\langle p_i\rangle$),} which respects our assumption that it is unknown which nodes are in the hidden community. 
We note that one could also define the ER null model with the observed mean edge probability 
\drt{$L\langle p_i\rangle  +  \sum_r \frac{K_r^2T_r}{N^2L} (\rho_r- \langle p_i\rangle)  $ 
to account for the slight increase in overall edge probability due to the presence of small communities. However, 
this change does not affect} the position of the dominant eigenvalues relative to the bulk, which is the relevant issue for community detectability \drt{as we will see below}. In particular, since \drt{$\frac{K_r^2T_r}{N^2L}\ll1$ for each $r$}, even the shift of the single associated eigenvalue within the bulk is negligible; therefore, we focus on the null model with expected edge weight $L\langle p_i\rangle$.

%

We develop random matrix theory based on the analysis in \cite{Benaych_2011,Nadakuditi2012}. To this end, we note that $ {\bf\overline{B}}$ can be written in the form
\begin{equation}
{\bf\overline{B}} = \langle {\bf\overline{B}} \rangle + {\bf X}, \label{eq:mod2}
\end{equation}
where 
\begin{align}
\langle {\bf\overline{B}} \rangle &= \sum_r \theta_r {\bf u}^{(r)}({\bf u}^{(r)})^T 
\end{align}
is a rank-$R$ matrix with eigenvalues given by
\begin{align}
\theta_r &= T_rK_r (\rho_r-\langle p_l\rangle),
\end{align}
and $\{{\bf u}^{(r)}\}$ are normalized indicator vectors for the $R$ communities that have entries 
\begin{align}
u_i^{(r)}=\left\{ \begin{array}{rl} \sqrt{1/K_r} ,&i\in\mathcal{K}_r \\ 0,& \text{otherwise.}\end{array}\right.
\end{align} 

The random matrix ${\bf X}$ has zero-mean entries $X_{ij}$ with variance $T\rho_r(1-\rho_r) + (L-T_r)\langle  p_l(1-p_l)\rangle$ if  $(i,j)\in\mathcal{K}_r\times\mathcal{K}_r$ and $L\langle  p_l(1-p_l)\rangle$ otherwise. 
In the $N\to\infty$ limit, and assuming the sizes $\{K_r\}$ grow more slowly than $N$, then the $\sum_r K_r^2 \ll N^2$ matrix entries corresponding to communities become negligible and $\bf X$ limits to a Wigner matrix \cite{Bai2010}. This allows us to use known results for the limiting dominant eigenvector of low-rank perturbations of Wigner matrices with variance $1/N$. Specifically, we define $\gamma = 1/ \sqrt{NL\langle  p_l(1-p_l)\rangle}$ so that the matrix $\gamma {\bf X}$ has entries with variance $1/N$ in the limit. We similarly define
\begin{equation}
\overline{\theta}_r = \gamma\theta_r= \frac{T_rK_r}{\sqrt{NL}}\frac{ \rho_r-\langle p_l\rangle}{\sqrt{\langle  p_l(1-p_l)\rangle}} \label{eq:theta}
\end{equation}
so that $\gamma\overline{{\bf B}}= \sum_r \overline{\theta}_r{\bf u}^{(r)}({\bf u}^{(r)})^T + \gamma{\bf X}$. It follows that the limiting $N\to\infty$ dominant eigenvectors $\{{\bf v}^{(r)}\}$ of $\gamma\overline{{\bf B}}$ (and of $\overline{{\bf B}}$, since scalar multiplication does not affect eigenvectors) satisfies \cite{Capitaine2009,Benaych_2011}
\begin{equation}
|\langle {\bf v}^{(r)},{\bf u}^{(r)}\rangle|^2 = \left\{
\begin{array}{rl}
1-1/\overline\theta^{2},&\overline\theta >1\\
0,&\text{otherwise}.
\end{array}
\right.
\label{eq:innerprod1}
\end{equation}
\drt{Note that we have assumed the dominant eigenvectors have been suitably enumerated so that $\mathbf{v}^{(r)}$ corresponds to the eigenvector localizing on community $r$.}
The value $\overline{\theta}_r=1$ identifies critical points at which there is a phase transition in eigenvector localization and detectability for community $r$, and this gives the critical community size
\begin{equation}
K_r^*  = \sqrt{T_r^{-2}NL} \frac{\sqrt{\langle  p_l(1-p_l)\rangle}}{ \rho_r-\langle p_l\rangle} . \label{eq:K}
\end{equation}
That is, a small community can be detected using a dominant vector ${\bf v}^{(r)}$ of $\overline{{\bf B}}$ only when $K_r>K_r^*$. We note that setting $L=T_r=1$, $\rho_r=1$ and $p_l=p$ in Eq.~\eqref{eq:K} recovers $K_r^*  = \sqrt{ {N p}/{( 1-  p )} }$, which describes the detectability transition for a single planted clique in a single-layer network \cite{Alon1998}.

We highlight \drt{an} important consequence of Eq.~\eqref{eq:K}. First, if the community persists across some fixed fraction of the layers, $T(L)=cL$, then $K_r^*\varpropto\sqrt{N/L}$; therefore, if $N$, $p$ and $T_r/L$ are held fixed and $L$ increases, then $K_r^*$ vanishes with scaling $O(L^{-1/2})$. \drt{This square-root scaling behavior is similar to that obtained for detection in  layer-aggregation  of large-scale communities that persist across all layers \cite{Taylor2016}.} Second, for fixed $N$ and $p$, a community of fixed size $K_r$ and persistence $T_r$ will become impossible to detect as $L$ increases, because $K^*_r$ increases with scaling $O(L^{1/2})$. This result highlights the importance of knowing which layers potentially contain the community, since \drt{the aggregation} of layers lacking the community can severely inhibit its detection.

\drt{Digging further, one can let $T_r$ vary with $L$ and then ask how $K_r^*$ depends on the scaling behavior for $T_r$. For $T_r \varpropto L^\beta$ Eq.~\eqref{eq:K} implies $K_r^*\varpropto L^{1/2-\beta}$ so that as $L\to\infty$,
\begin{align}\label{eq:beta}
K_r^* &\to \left\{ \begin{array}{rl} 0, & \beta >1/2 \\ \infty,  & \beta <1/2. \end{array}\right.
\end{align}
That is, $T_r$, the number of layers containing the community, must increase with $L$ at least as $\mathcal{O}(L^{1/2})$, otherwise summation-based layer aggregating will inhibit (rather than promote) small-community detection. Note that $T\varpropto L^{-1/2}$ is a critical case in which $K_r^*$ is independent of $L$. We highlight that Eq.~\eqref{eq:beta} is somewhat surprising since
summation-based aggregation benefits detection even if the fraction $T_r/L$ of layers containing the community vanishes with  $L$, provided that it decays more slowly than $\mathcal{O}(L^{-1/2})$.}

\begin{figure}[t]
\centering
\epsfig{file = 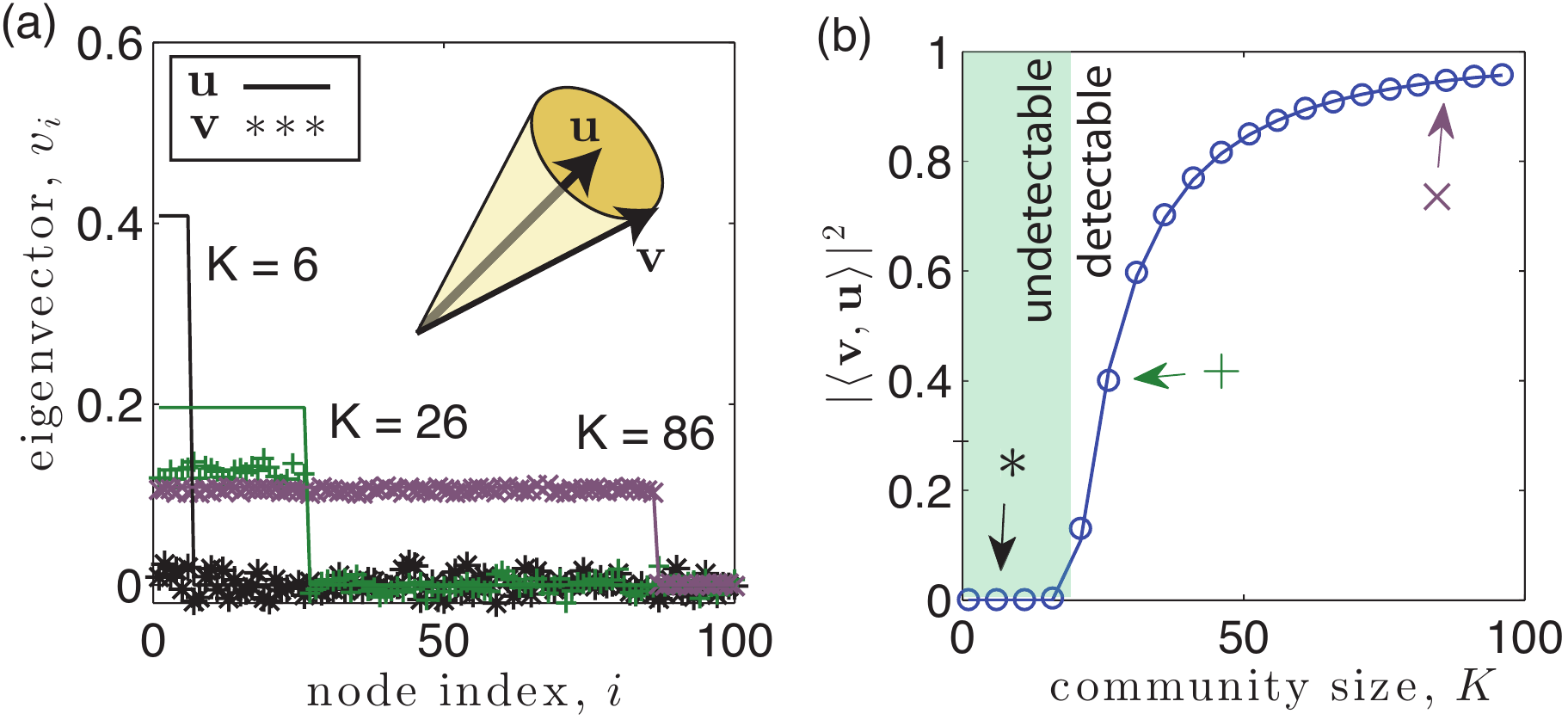, clip =,width=1\linewidth}
\caption{{\it Eigenvector localization yields detectability phase transition.}
(a)~Entries $v_i^{(r)}$ (symbols) of a dominant eigenvector of the modularity matrix for the summation network of a multilayer network with a hidden community of size $K_r$. Parameters include $T_r=2$, $L=16$, $N=10^4$, $\rho=1$ and the edge probabilities $\{p_l\}$ of layers are Gaussian distributed with mean $\langle p_l\rangle =0.01$ and standard deviation $\sigma_p=0.001$.
To allow visualization, we assume nodes $i\in\{1,\dots,K\}$ are in the community, and we only visualize $v_i^{(r)}$ for nodes $i\in\{1,100\}$. As shown by the illustration, as $K_r$ increases ${\bf v}^{(r)}$ aligns with the indicator vector ${\bf u}^{(r)}$, which is nonzero only for the $K_r\ll N$ entries $u_i^{(r)}$ that correspond to nodes in the community, $\mathcal{K}_r$.
(b)~Observed (symbols) and predicted (curves) values of $|\langle {\bf v}^{(r)},{\bf u}^{(r)} \rangle |^2$ given by Eq.~\eqref{eq:innerprod1} quantify this localization phenomenon. Arrows indicate the values of $K$ used for panel (a). The critical size $K_r^*$ such that $|\langle {\bf v}^{(r)},{\bf u}^{(r)} \rangle |^2=0$ for $K_r\le K_r^*$, whereas $|\langle {\bf v}^{(r)},{\bf u}^{(r)} \rangle |^2>0$ for $K_r>K_r^*$ marks a phase transition---that is, both in terms of eigenvector localization and detectability of the community.
} 
\label{fig:Localization}
\end{figure}

\begin{figure}[t]
\centering 
\epsfig{file = 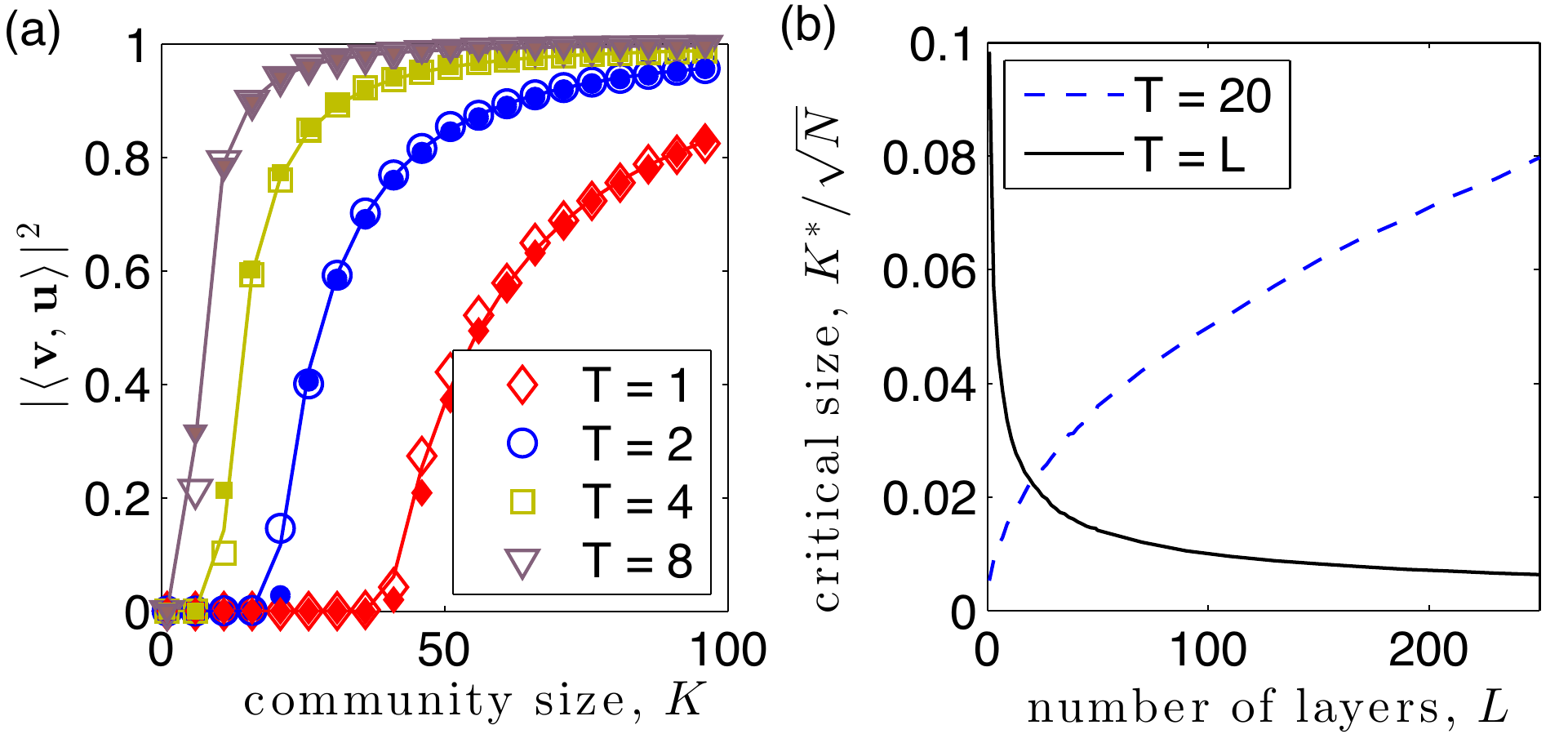, clip =,width=1\linewidth } 
\caption{{\it Influence of community persistence $T_r$ on eigenvector localization for  \drt{summation-based layer aggregation}.}
(a)~Observed (symbols) and predicted values of $|\langle {\bf v}^{(r)},{\bf u}^{(r)} \rangle |^2$ given by Eq.~\eqref{eq:innerprod1} (curves) versus $K_r$ for $T_r\in\{1,2,4,8\}$. Open symbols indicate the parameters used in Fig.~\ref{fig:Localization}, whereas filled symbols indicate when the layers' edge probabilities $\{p_l\}$ are drawn uniformly from $[0,0.02]$, and we plot the mean value of $|\langle {\bf v}^{(r)},{\bf u}^{(r)} \rangle |^2$ across 10 choices for the sets $\mathcal{K}_r$ and $\mathcal{T}_r$.
(b)~Critical size $K_r^*$ given by Eq.~\eqref{eq:K} versus $L$ for fixed $T_r$ (dashed) and $T_r=L$ (solid). \drt{As indicated by Eq.~\eqref{eq:beta}, layer aggregation by summation can enhance or inhibit detection depending on whether or not the scaling for $T_r(L)$ exceeds $  \mathcal{O}(L^{1/2})$.}
} 
\label{fig:summation_depend}
\end{figure}

 \subsubsection{Numerical Validation and Scaling Behavior}\label{sec:detect_sum}


We support Eqs.~\eqref{eq:innerprod1}--\eqref{eq:K} in Fig.~\ref{fig:Localization}, using numerical experiments with $N=10^4$ nodes and edge probabilities $\{p_l\}$ drawn from a Gaussian distribution with mean $p=0.01$ and standard deviation $\sigma_p=0.001$. We focus on the case of clique detection (i.e., $\rho=1$), hiding the clique in $T=2$ of the $L=16$ layers.
In Fig.~\ref{fig:Localization}(a), we plot the entries $\{v_i^{(r)}\}$ (symbols) of the dominant eigenvector of the modularity matrix for the summation network as well as the entries $\{u_i^{(r)}\}$ for the indicator vector, which are nonzero only for nodes $i\in\mathcal{K}$ involved in the clique. We show results for community sizes $K_r\in\{6,26,86\}$, which respectively place the system below, just above, and well above the phase transition. The illustration highlights that as $K$ increases, vector ${\bf v}^{(r)}$ aligns with ${\bf u}^{(r)}$. We quantify this localization phenomenon by plotting in Fig.~\ref{fig:Localization}(b) observed (symbols) and predicted values of $|\langle v,u\rangle|^2$ given by Eq.~\eqref{eq:innerprod1} (curve). Note that the values of $|\langle v^{(r)},u^{(r)}\rangle|^2$ depict a phase transition that occurs at a critical subgraph size $K_r^*$ given by Eq.~\eqref{eq:K}: $|\langle v^{(r)},u^{(r)}\rangle|^2>0$ when $K_r>K_r^*$, whereas $|\langle v,u\rangle|^2=0$ when $K_r\le K_r^*$. This phase transition in eigenvector localization drives a phase transition for community detection based on ${\bf v}^{(r)}$. Arrows indicate the values of $K_r$ used in panel (a).

In Fig.~\ref{fig:summation_depend}(a), we compare observed (symbols) and predicted values of $|\langle v,u\rangle|^2$ given by Eq.~\eqref{eq:innerprod1} (curves) for varying $K_r$ with $T_r\in\{1,2,4,8\}$. Open symbols indicate the parameters used in Fig.~\ref{fig:Localization}, whereas filled symbols indicate the mean value of $|\langle {\bf v},{\bf u} \rangle |^2$ for 10 trials in which the layers' edge probabilities $\{p_l\}$ are drawn uniformly from $[0,0.02]$. Note that as $T_r$ increases, the curves shift to the left, illustrating that as the community persists across more layers, the localization phenomenon is stronger and the hidden community is easier to detect. 
In Fig.~\ref{fig:summation_depend}(b), we study the dependence of $K_r^*$ on the number of layers, $L$, and we compare the effect of keeping $T_r$ fixed versus allowing $T_r$ to grow with $L$. Specifically, we set either $T_r(L)=20$ or $T_r(L)=L$, and we plot the value of $K_r^*$ given by Eq.~\eqref{eq:K}. Note that if the community persists across a fraction of the layers---that is, $T_r(L)=cL$ for some constant c---then $K_r^*$ vanishes with scaling $O(L^{-1/2})$. However, if $T_r$ is held fixed, then $K_r^*$ increases with scaling $O(L^{1/2})$.

\drt{In summary, these experiments illustrate how layer aggregation through summation can enhance small-community detection if the community persists across sufficiently many 
layers, but it can obscure detection 
if the community is present in too few layers.
We will see in the next section that thresholding the summation can help overcome this problem, potentially reducing the detectability limit  by orders of magnitude to yield super-resolution community detection.}

\subsection{Thresholding as a Nonlinear Data Filter}\label{sec:thresh}

\subsubsection{Random Matrix Theory for Modularity Matrices}

\drt{
We now study layer aggregation with thresholding as a nonlinear data filter that enhances small-community detection.
We begin by solving for \emph{effective} edge probabilities for the thresholding process} \cite{Taylor2016}. Thresholding the summation $\sum_l {\bf A}^{(l)}$ at $\tilde{L}$ yields a binary adjacency matrix $\hat{{\bf A}}^{(\tilde{L})}$ with entries $\hat{A}^{(\tilde{L})}_{ij}\in\{0,1\}$ indicating whether or not $\overline{A}_{ij}\ge \tilde{L}$. 
\drt{For  edges $(i,j)\not \in \cup_r \{\mathcal{K}_r\times \mathcal{K}_r\}$,}
$\overline{A}_{ij}$ follows a Poisson binomial distribution $f_{PB}(a;L,\{p_l\})$ given by Eq.~\eqref{eq:Entry_dist} and the inequality is satisfied with probability 
\begin{align}
\hat{p}^{(\tilde{L})}&=P\left[\overline{A}_{ij}\ge \tilde{L}\right]
=
1- F_{PB}(\tilde{L}-1,L,\{p_l\}),\label{eq:phat}
\end{align}
where $F_{PB}(a,L,\{p_l\})$ is the associated cumulative distribution function (CDF). 
\drt{For  edges $(i,j)\in\{\mathcal{K}_r\times\mathcal{K}_r\}$,}
\drt{$\overline{A}_{ij}$ follows a Poisson binomial distribution $f_{PB}(a;L,\{q^{(r)}_l\})$ given by Eq.~\eqref{eq:Entry_dist} and the inequality is satisfied with probability 
\begin{align}
\hat{\rho}_r^{(\tilde{L})}
&=P\left[\overline{A}_{ij}\ge \tilde{L}\right]
=&1- F_{PB}(\tilde{L}-1,L,\{q^{(r)}_l\}),\label{eq:rhohat}
\end{align}
where $q^{(r)}_l=\rho_r$ for $l\in\mathcal{T}_r$ and otherwise $q^{(r)}_l=p_l$.}
%
In the case of a clique (i.e., $\rho_r=1$), Eq.~\eqref{eq:rhohat} \drt{can be written as} 
\begin{equation}\label{eq:rhohat_clique}
\hat{\rho}_r^{(\tilde{L})}= 1- F_{PB}(\tilde{L}-T_r-1,L-T_r,\{p_l\}_{l\not\in \mathcal{T}_r}).
\end{equation}

Given the effective edge probabilities for the network and a community (i.e., $\hat{p}^{(\tilde{L})}$ and $\hat{\rho}_r^{(\tilde{L})}$, respectively), it is straightforward to study the detectability \drt{limits} of a \drt{community} for thresholded networks using Eqs.~\eqref{eq:innerprod1} and \eqref{eq:K}. In particular, we substitute $L=T_r=1$ to obtain
\begin{equation}
|\langle{\bf \hat{v}}^{(r)}, {\bf u}^{(r)}\rangle|^2 = \left\{
\begin{array}{rl}
1-1/\hat{\theta}_r^{2},&\hat{\theta}_r >1\\
0,&\text{otherwise},
\end{array}
\right.
\label{eq:innerprod2}
\end{equation}
where ${\bf\hat{v}}^{(r)}$ is a dominant eigenvector of modularity matrix 
\begin{align}
\hat{{\bf B}}=\hat{{\bf A}}^{(\tilde{L})} - \hat{p}^{(\tilde{L})}{\bf 1}{\bf 1}^T \label{eq:Bhat}
\end{align} 
and 
$\hat{\theta}_r = K(\hat{\rho}_r^{(\tilde{L})}- \hat{p}^{(\tilde{L})})/ \sqrt{N  \hat{p}^{(\tilde{L})}(1-\hat{p}^{(\tilde{L})}) }.$ 
Setting $\hat{\theta}_r=1$ gives a detectability limit \drt{for each community $r$} in terms of the effective edge probabilities $\hat{p}^{(\tilde{L})}$ and $\hat{\rho}_r^{(\tilde{L})}$,
\begin{equation}
\hat{K}_r^* =  \frac{\sqrt{N \hat{p}^{(\tilde{L})}( 1-  \hat{p}^{(\tilde{L})} )}}{  \hat{\rho}_r^{(\tilde{L})}-\hat{p}^{(\tilde{L})} } .\label{eq:Khat}
\end{equation}
Equations~\eqref{eq:innerprod2}--\eqref{eq:Khat} illustrate that the detectability \drt{limits} for  thresholded networks depend only on the effective edge probabilities; however, these depend sensitively on the choice of threshold $\tilde{L}$.

\drt{Importantly,  $\hat{K}_r^*$ given by Eq.~\eqref{eq:Khat} can potentially be orders of magnitude smaller than ${K}_r^*$ given by Eq.~\eqref{eq:K}, a phenomenon we call super-resolution detection.
In addition to numerical experiments that will follow below, we further study this phenomenon by comparing  $\hat{K}_r^*$ and ${K}_r^*$ for network parameters wherein we can obtain deeper insight. We consider clique detection (i.e., $\rho_r=1$) in a sparse network (i.e.,  $p_l\ll 1$) and focus on the threshold value $\tilde{L}=T_r$ to obtain  
\begin{equation}
\hat{K}_r^* \approx   \sqrt{N}    \sqrt{ \hat{p}^{(T_r)}}  .\label{eq:Khat_proof}
\end{equation}
Using these assumptions also in Eqs.~\eqref{eq:phat} and \eqref{eq:rhohat_clique}, we find the effective edge probabilities 
$\hat{p}^{(T_r)} = 1- F_{PB}(T_r-1,L,\{p_l\}) $ and 
$\hat{\rho}_r^{(T_r)} =1$.
Furthermore, we apply Hoeffding's inequality \cite{Hoeffding} to obtain 
$\hat{p}^{(T_r)}   \le e^{-2L(\langle p_l \rangle -T_r/L)^2}.$
Noting $0 < \langle p_l \rangle \ll  T_r/L$, we find the $\langle p_l \rangle\to0$ limiting bound  
\begin{align}
\hat{p}^{(T_r)}  & \le e^{-2T_r^2/L},
\end{align}
illustrating that $\hat{p}^{(T_r)}$ and $ \hat{K}_r^*$ decay exponentially with $T_r^2/L$.
%
%
On the other hand,
we use the sparsity assumption in Eq.~\eqref{eq:K} to obtain
\begin{equation}
K_r^*  \approx \frac{\sqrt{N L\langle  p_l\rangle}}{\sqrt{T_r^2}} . \label{eq:K_proof}
\end{equation}
Thus, in this case $K_r^* $ decays  as  $\mathcal{O}(1/\sqrt{T_r^2/L})$, whereas $\hat{K}_r^*$ decays exponentially (i.e., considerably faster) with $T_r^2/L$. 
}


\subsubsection{Numerical Validation and Super-Resolution Detection}\label{sec:detect_thresh}

\begin{figure}[t]
\centering
\epsfig{file = 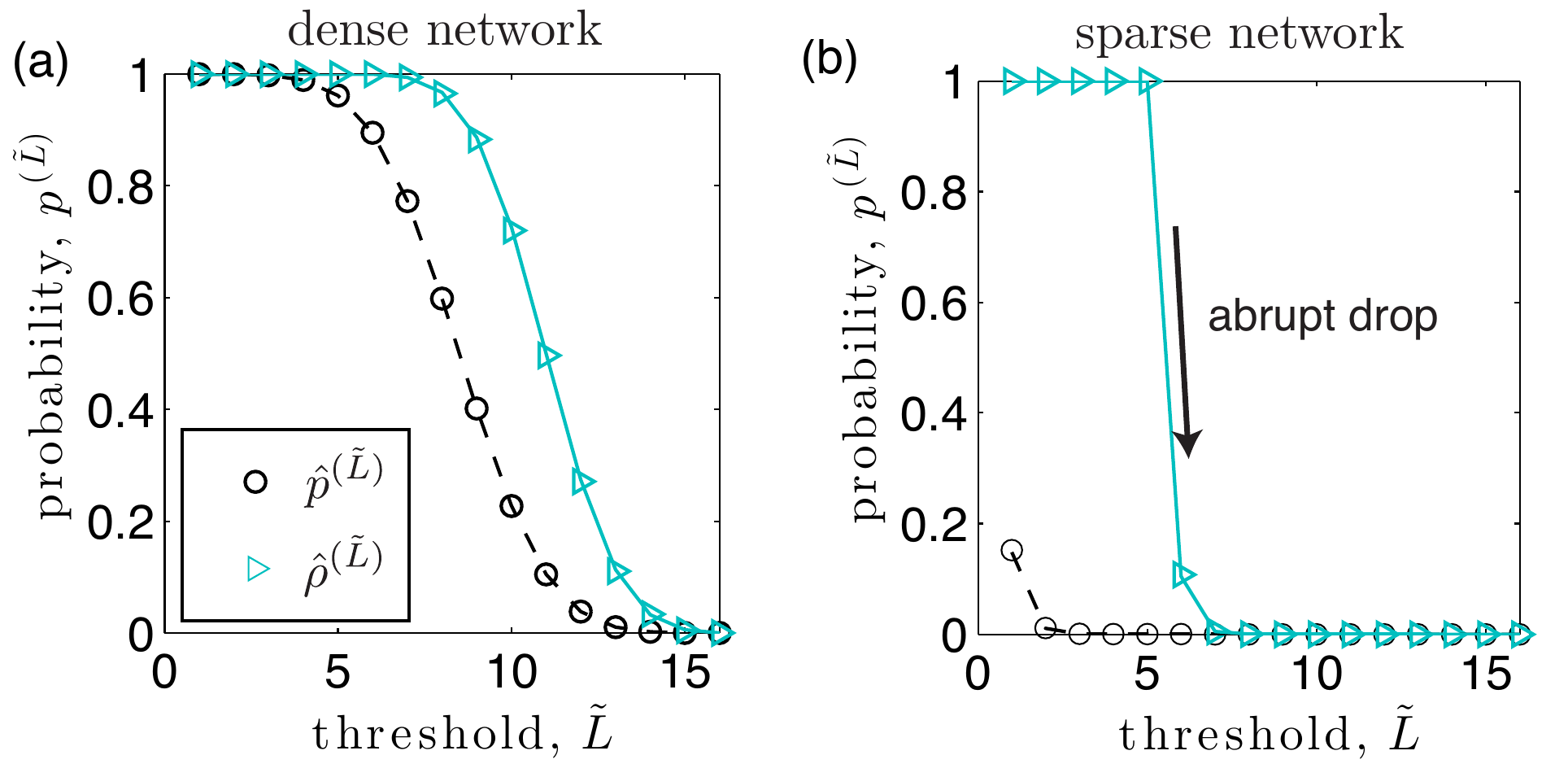, clip =,width=1\linewidth }
\caption{{\it Effective edge probabilities for threshold-based layer aggregation.}
Observed (symbols) and predicted values given by Eqs.~\eqref{eq:phat} and \eqref{eq:rhohat_clique} (curves) for the effective edge probability of the background network, $\hat{p}^{(\tilde{L})}$, and for a community, $\hat{\rho}_r^{(\tilde{L})}$, as a function of $\tilde{L}$. Network parameters include $N=10^4$, $L=16$, $T=5$, and $\sigma_p=0.001$ and either (a) $\langle p_l \rangle=0.5$ or (b) $\langle p_l \rangle=0.01$. Note for the sparse network in panel (b) that $\hat{\rho}^{(\tilde{L})}$ undergoes an abrupt drop when $\tilde{L}$ surpasses  $ T_r=5$.
} 
\label{fig:Thresholded}
\end{figure}
\begin{figure}[t]
\centering
\epsfig{file = 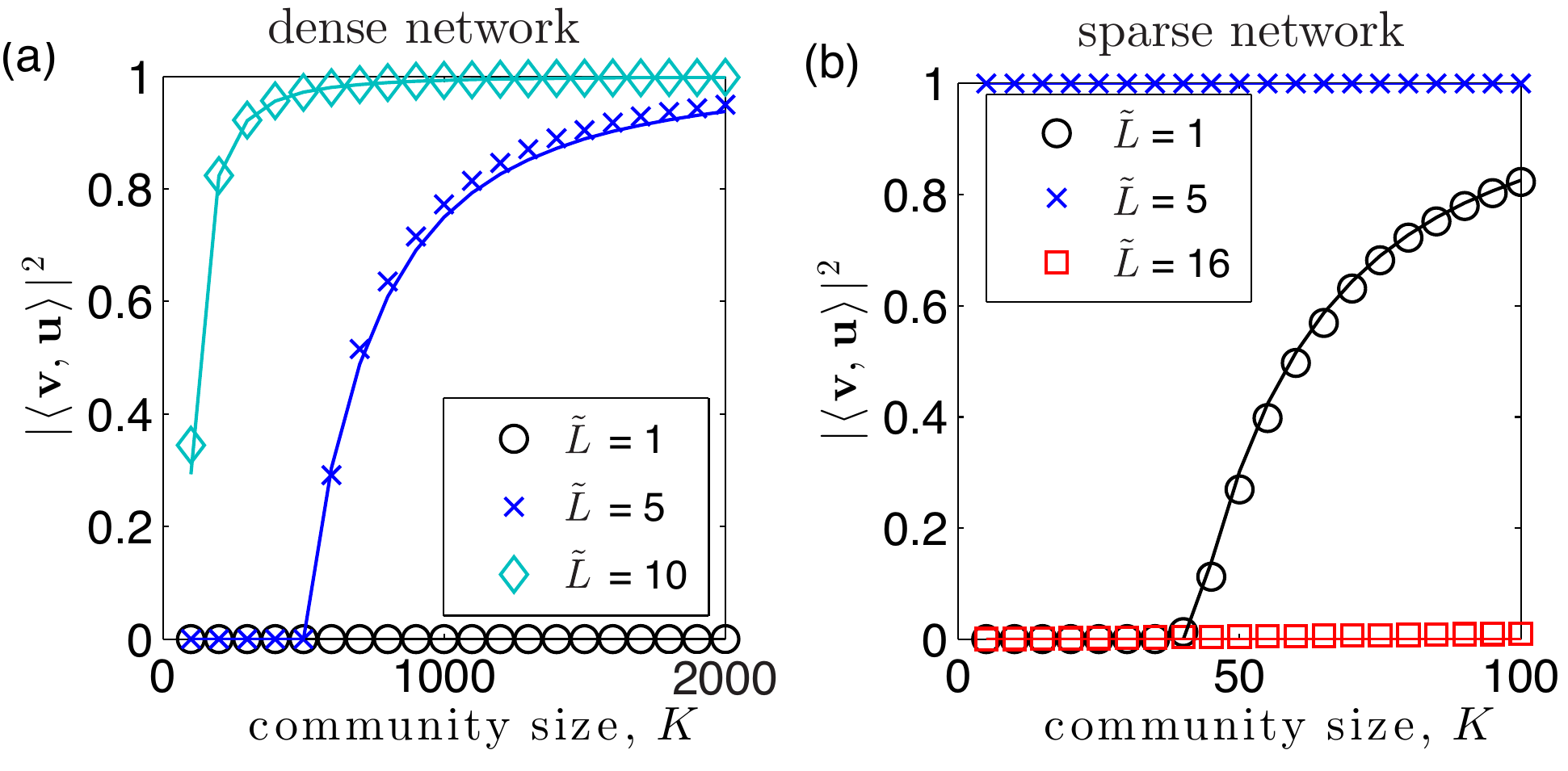, clip =,width=1\linewidth } 
\caption{{\it Detectability phase transitions for threshold-based layer aggregation.}
We plot $|\langle {\bf v}^{(r)},{\bf u}^{(r)} \rangle |^2$ versus community size $K_r$ with identical parameters to those used to produce Fig.~\ref{fig:Thresholded} except with selected choices for the threshold $\tilde{L}$. 
} 
\label{fig:Thresholded2}
\end{figure}

We now support Eqs.~\eqref{eq:phat}--\eqref{eq:Khat} with numerical experiments \drt{and illustrate that certain thresholds  lead to super-resolution community detection}. We consider the detection of a dense subgraph that is hidden in both (a) a dense network with $\langle p_l \rangle=0.5$ and (b) a sparse network with $\langle p_l \rangle=0.01$. Both networks were constructed with $N=10^4$, $\sigma_p=0.001$, $\rho_r=1$, $L=16$, and $T_r=5$. 

In Fig.~\ref{fig:Thresholded}, we compare observed (symbols) and predicted values (curves) of the effective edge probabilities $\hat{p}^{(\tilde{L})}$ given by Eq.~\eqref{eq:phat} and $\hat{\rho}_r^{(\tilde{L})}$ given by Eq.~\eqref{eq:rhohat} as a function of the threshold $\tilde{L}$. Note in both panels that the effective edge probability $\hat{p}^{(\tilde{L})}$ of the background network always decays with increasing $\tilde{L}$.
In contrast, the effective edge probability between nodes in the community depends on whether or not $\tilde{L}>T_r$:
$\hat{\rho}_r^{(\tilde{L})}=1$ when $\tilde{L}\le T_r$ since $\rho=1$, whereas $\hat{\rho}_r^{(\tilde{L})}$ decays with increasing $\tilde{L}$ for $\tilde{L}>T_r$ . Importantly, the rate of decay depends on the network's mean edge density $\langle p_l\rangle$:  $\hat{\rho}^{(\tilde{L})}$ slowly decreases for the dense network, whereas it abruptly drops for the sparse network. 

In Fig.~\ref{fig:Thresholded2}, we plot observed (symbols) and predicted values (curves) for $|\langle {\bf v}^{(r)},{\bf u}^{(r)} \rangle |^2$ given by Eq.~\eqref{eq:innerprod2} versus $K$ for different choices of $\tilde{L}$. The parameters used are identical to those of Fig.~\ref{fig:Thresholded} and panels (a) and (b) again depict results for $\langle p_l \rangle=0.5$ and $\langle p_l \rangle=0.01$, respectively. 
We highlight several important observations. 
First, note in both panels that $\tilde{L}=T_r=5$ yields better detectability than $\tilde{L}=1$. However, when $\tilde{L}>   T_r$ we find contrasting results for sparse and dense networks. For the sparse network shown in Fig.~\ref{fig:Thresholded2}(b), the hidden community becomes harder to detect when $\tilde{L}>  T_r$ (see curve for $\tilde{L}=16$), which intuitively occurs because $\hat{\rho}_r^{(\tilde{L})}$ rapidly decays and the thresholded networks will no longer contain a dense subgraph. On the other hand, for the dense network depicted in Fig.~\ref{fig:Thresholded2}(a), increasing $\tilde{L}$ can improve detectability when $\tilde{L}> T_r$ (see curve for $\tilde{L}=10$).

\begin{figure}[t]
\centering
\epsfig{file = 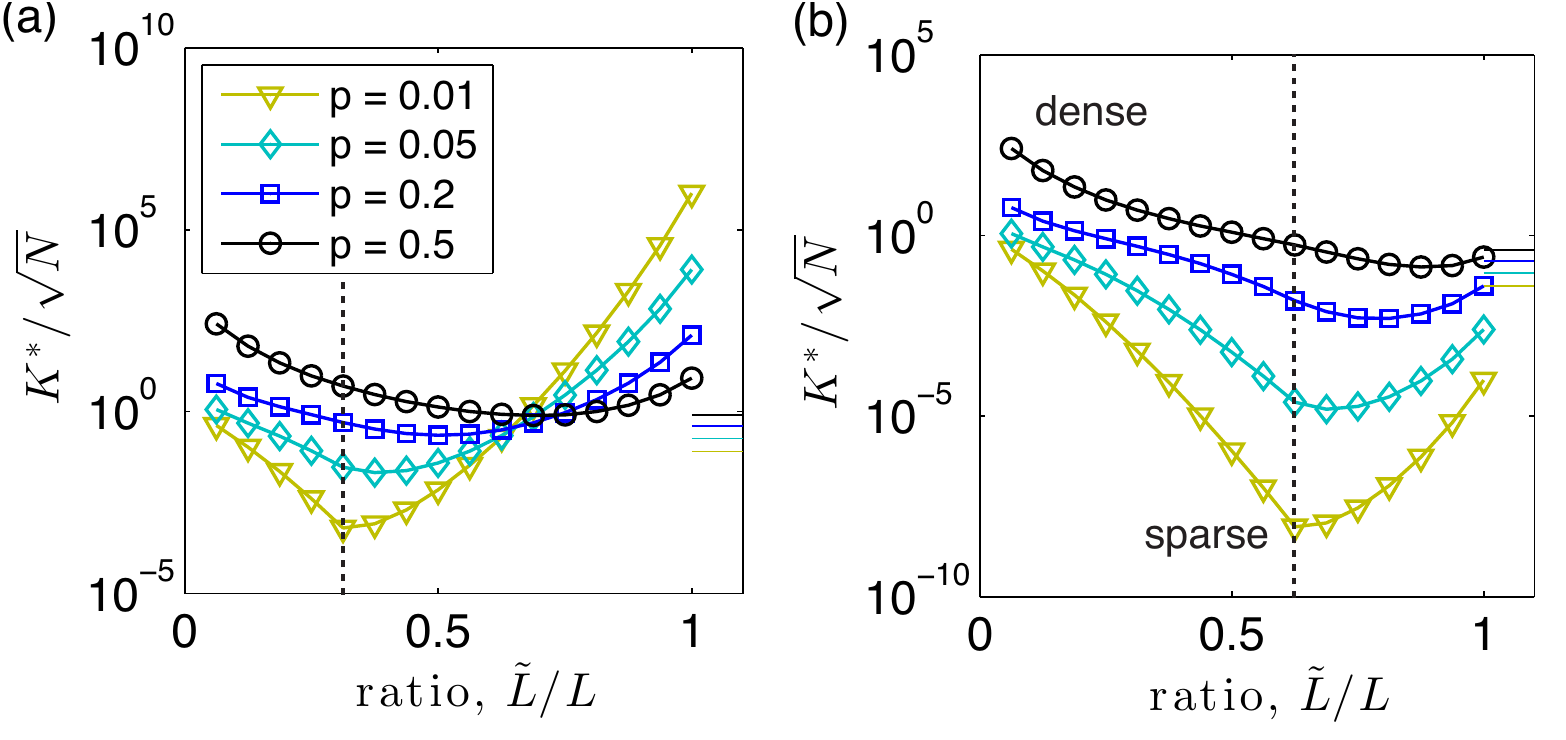, clip =,width=1\linewidth }
\caption{{\it \drt{Super-resolution community detection for threshold-based layer aggregation.}}
We plot $\hat{K}_r^*$ given by Eq.~\eqref{eq:Khat} as a function of $\tilde{L}$ for $p \in\{0.01,0.05,0.2,0.5\}$, $N=10^4$, $\rho=1$, $\sigma_p=0.001$, $L=16$ and either (a) $T_r=5$ or (b) $T_r=10$. Note that the $\tilde{L}$ value yielding the minimum $\hat{K}_r^*$ occurs at $\tilde{L}=  T_r$ (vertical dotted lines) for sparse networks, whereas it increases with increasing $p$ (e.g., compare $p=0.01$ and $p=0.5$ in panel b). 
The horizontal lines on the right edge of the panels indicate $K_r^*$ given by Eq.~\eqref{eq:K} for summation networks. 
\drt{Importantly, thresholding can potentially decrease $\hat{K}_r^*$ by many orders of magnitude as compared to $K_r^*$.}
} 
\label{fig:Thresholded3}
\end{figure}

\begin{figure*}[t]
\centering
\epsfig{file = 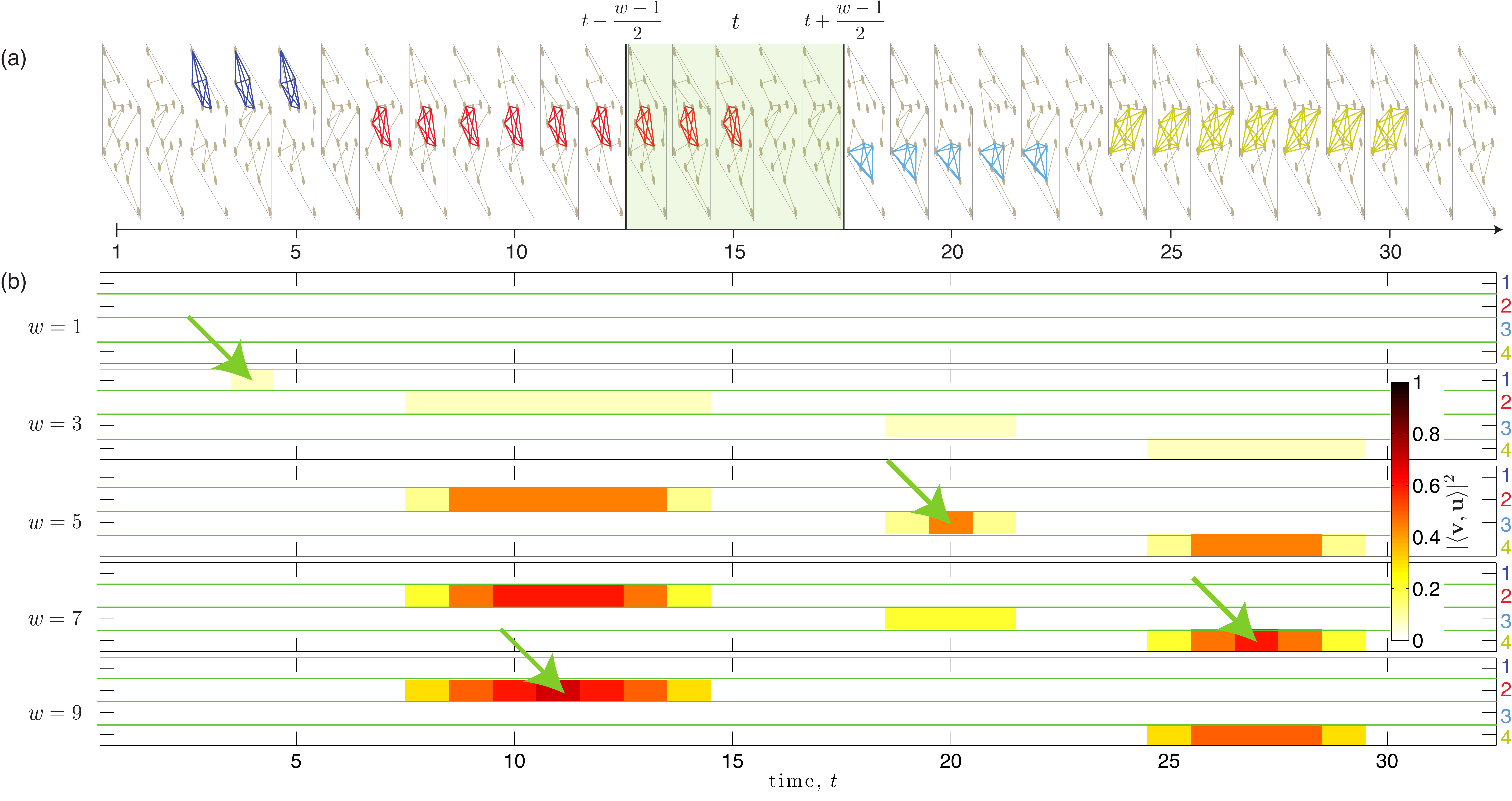, clip =,width=1\linewidth }
\caption{{\it Detectability of small communities in temporal networks with summation-based binning into time-windows.}
(a)~Illustration of a temporal network with $L=32$ time layers and hidden communities that persist across different time layers. The shaded region indicates a bin, or time window, of size $w\le L$ at time $t$ for which the layers will be aggregated, which is a process that can be used to discretize and/or smooth the network data. The bin contains layers $\mathcal{W}_w(t)=\{t-(w-1)/2,\dots,t+(w-1)/2\}$.
(b)~We illustrate by color the values $|\langle {\bf v}^{(r)},{\bf u}^{(r)} \rangle |^2$ for the aggregation of layers across bins $\mathcal{W}_w(t)$ for each of the four communities $r\in\{1,2,3,4\}$. In particular, we show Eq.~\eqref{eq:innerprod1} under the variable substitutions $T_r(\mathcal{W}_w(t)) \mapsto T$ and $w\mapsto L$, where $T_r(\mathcal{W}_w(t))$ is the number of layers in which community $r$ is present in bin $\mathcal{W}_w(t)$. Layer aggregation across each bin was implemented by summation. We study a temporal network with $N=10^4$, $L=32$, $p=0.01$, $\sigma_p=0.001$, and we show results for several bin widths $w\in\{1,3,5,7,9\}$ The hidden communities all contain $K_r=8$ nodes and have different persistent lengths $T_r$ as depicted in panel (a).  The green arrows indicate for each $r$ the bin location and $w$ value at which $|\langle {\bf v}^{(r)},{\bf u}^{(r)} \rangle |^2$ obtains its maximum. 
%
} 
\label{fig:Time_pic}
\end{figure*}


\drt{We now present an experiment highlighting the occurrence of super-resolution community detection for certain threshold values.} In Fig.~\ref{fig:Thresholded3}, we study the dependence of the critical community size $K_r^*$ on the threshold $\tilde{L}$. We plot $\hat{K}_r^*$ given by Eq.~\eqref{eq:Khat} as a function of $\tilde{L}$ for $p \in\{0.01,0.05,0.2,0.5\}$, $N=10^4$, $\rho=1$, $\sigma_p=0.001$, $L=16$ and either (a) $T_r=5$ or (b) $T_r=10$. Note for the sparsest network, i.e., $p=0.01$, that the minimum value of $K^*$ occurs when $\tilde{L}=  T_r$ (vertical dashed line). Interestingly, as the mean edge density $p=\langle p_l\rangle$ increases, the threshold $\tilde{L}$ at which $\hat{K}_r^*$ attains its minimum value shifts from $\tilde{L}= T_r$ towards $\tilde{L}=L$. The horizontal lines on the right edge of the panels indicate $K_r^*$ given by Eq.~\eqref{eq:K} for the summation network.  

\drt{Importantly, note that} for a wide range of parameters  $\hat{K}_r^*$ for the thresholded networks is significantly smaller than $K_r^*$ for the corresponding summation networks. 
\drt{In particular, one can observe for $p=0.1$ and $\tilde{L}/L=T_r/L$ in Fig.~\ref{fig:Thresholded3}(b) that $\hat{K}_r^*$ is  many orders of magnitude smaller than $K_r^*$ [$\mathcal{O}(10^{-6})$ times here] .}
That is, thresholding the summation \drt{can dramatically} improve detectability as compared to summation without thresholding. This surprising result 
contrasts our previous findings for the detectability of large communities that persist across all layers \cite{Taylor2016}, where it was found that thresholding always inhibited detection \drt{(although optimal thresholds were found to minimize inhibition)}.

\begin{figure*}[t]
\centering
\epsfig{file = 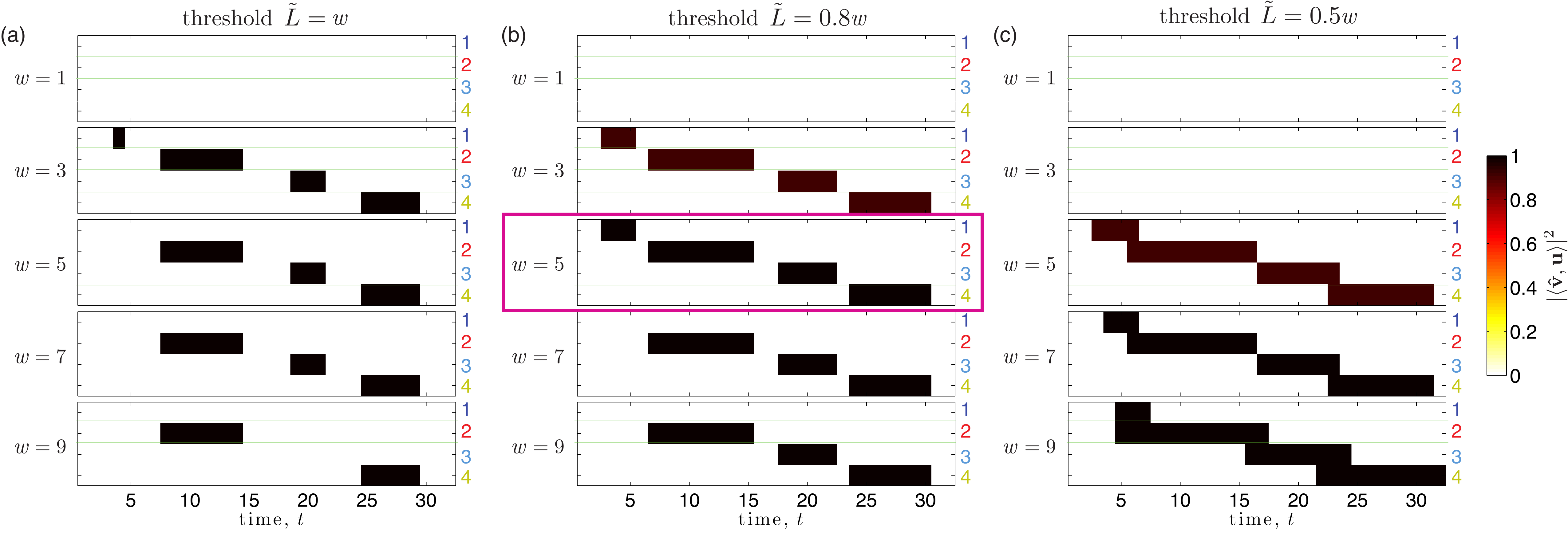, clip =,width=1\linewidth }
\caption{{\it Detectability of small communities in temporal networks with time-window binning by summation and thresholding.}
We illustrate by color the values $|\langle {\bf \hat{v}}^{(r)},{\bf u}^{(r)} \rangle |^2$ given by Eq.~\eqref{eq:innerprod2} for each of the four communities $r\in \{1,2,3,4\}$ with the variable substitutions $T_r(\mathcal{W}_w(t)) \mapsto T$ and $w\mapsto L$ into Eqs.~\eqref{eq:phat}--\eqref{eq:Khat}. Results are shown for bins of width $w\in\{1,3,5,7,9\}$ for a temporal network with $N=10^4$ nodes, $L=32$ time layers, and hidden communities as depicted in Fig.~\ref{fig:Time_pic}(a). The communities \drt{each} contain $K_r=K=8$ nodes and have different persistence lengths $T_r$. Layer aggregation across each bin was implemented by summation and thresholding at $\tilde{L}$. Panels (a), (b) and (c) respectively indicate the choices $\tilde{L}=w$, $\tilde{L}=0.8w$ and $\tilde{L}=0.5w$. 
\drt{The violet box in panel (b) indicates combinations of thresholds and bin sizes that yield accurate detection of all four communities. 
We stress, however, that since the the detectability-limit criterion given by Eq.~\eqref{eq:Khat} depends on a complex interplay between the community and network characteristics, one should not in general expect there to exist a \emph{single} best combination for \emph{all} communities.}
} 
\label{fig:Time_pic2}
\end{figure*}

\section{Small-community detection in time-varying networks}\label{sec:time}
%
We now present an experiment involving small-community detection in time-varying networks to highlight several practical insights following from our theoretical results.
%
%
%
%
Note that unlike Sec.~\ref{sec:analysis}, where there were no restrictions on which layers a community persists, we now assume that each community persists across consecutive layers.
We conducted experiments for a synthetic temporal network with $N=10^4$ nodes and $L=32$ time layers, each of which is drawn from an ER network with edge probability $p_l$, which we drew from a Gaussian distribution with mean $p=0.01$ and standard deviation $\sigma_p=0.001$. We then planted $R=4$ communities, each involving $K_r=K=8$ nodes, in the following sets of layers: $\mathcal{T}_1=\{3,4,5\}$ for community 1, $\mathcal{T}_2=\{7,\dots,15\}$ for community 2, $\mathcal{T}_3=\{18,\dots,22\}$ for community 3, and $\mathcal{T}_4=\{24,\dots,30\}$ for community 4. 
In Fig.~\ref{fig:Time_pic}(a), we provide a representative illustration of the temporal network, where we indicate in which layers the  communities are present. We also illustrate by the shaded region an example time window, or bin, $\mathcal{W}_w(t)=\{t-(w-1)/2,\dots,t+(w-1)/2\}$ for $t\in\{(w-1)/2,L-(w-1)/2\}$ that contains layers to be aggregated. 

%
We first consider aggregation by summation. In Fig.~\ref{fig:Time_pic}(b), we illustrate by color the values $|\langle {\bf v}^{(r)},{\bf u}^{(r)} \rangle |^2$ for the aggregation of layers across bins $\mathcal{W}_w(t)$. In particular, we show Eq.~\eqref{eq:innerprod1} under the variable substitutions $T_r(\mathcal{W}_w(t)) \mapsto T$ and $w\mapsto L$, where $T_r(\mathcal{W}_w(t))=|\mathcal{W}_w(t) \cap \mathcal{T}_r|$ is the number of layers in which community $r$ is present in bin $\mathcal{W}_w(t)$. We show results for several bin widths $w\in\{1,3,5,7,9\}$. 
The green arrows indicate for each $r$ the bin location and $w$ value at which $|\langle {\bf v}^{(r)},{\bf u}^{(r)} \rangle |^2$ obtains its maximum. As expected, $|\langle {\bf v}^{(r)},{\bf u}^{(r)} \rangle |^2$ obtains its maximum for each community $r$ when the bin $\mathcal{W}_w(t)$ is exactly the set of layers in which community $r$ is present, $\mathcal{W}_w(t)=\mathcal{T}_r$ (i.e., when $T_r = w$). 


Before studying aggregation by summation and thresholding, we first make several important observations using Fig.~\eqref{fig:Time_pic}. First, note for $w=1$ in panel (b) that no communities are detectable. In other words, all communities are undetectable if the layers are studied in isolation. However, they can be detected if the layers are binned into time windows. Second, because the optimal bin size $w$ is unique to every community (i.e., because they have different persistences $T_r\in[3,9]$), there is no bin size that is best for all communities. In fact, detectability requires $K_r>K_r^*$ given by Eq.~\eqref{eq:K}, which requires for each community that $w$ is not too large or too small. For example, community 1 is only detectable when $w=3$  and community 3 is only detectable when $w\in[3,7]$. 

\drt{One final important observation for Fig.~\ref{fig:Time_pic}(b) is that even} when communities are detectable, the values $|\langle {\bf v}^{(r)},{\bf u}^{(r)} \rangle |^2$ are not very large---specifically, $|\langle {\bf v}^{(r)},{\bf u}^{(r)} \rangle |^2\le0.7$ in all cases. \drt{This can be problematic since detection error rates increase as $|\langle {\bf v}^{(r)},{\bf u}^{(r)} \rangle |^2$ decreases, approaching 100\% error as $|\langle {\bf v}^{(r)},{\bf u}^{(r)} \rangle |^2\to0$. (See \cite{Nadakuditi2012} for an analysis of error rates based on a hypothesis-testing framework for clique detection in single-layer networks.)
Because $|\langle {\bf v}^{(r)},{\bf u}^{(r)} \rangle |^2$ remains small for community 1 for all choices of $w$, it effectively remains undetectable to summation-based layer aggregation.
}
 
\drt{We now illustrate layer aggregation with thresholding as a nonlinear data filter that can allow greatly improved small-community detection for the temporal network shown in Fig.~\ref{fig:Time_pic}(a), including the accurate recovery of community 1.} In Fig.~\ref{fig:Time_pic2}, we plot $|\langle {\bf \hat{v}}^{(r)},{\bf { u}}^{(r)} \rangle |^2$ given by Eq.~\eqref{eq:innerprod2} with the variable substitutions $T_r(\mathcal{W}_w(t)) \mapsto T$ and $w\mapsto L$ into Eqs.~\eqref{eq:phat}--\eqref{eq:Khat}. Results reflect the aggregation of layers into bins $\mathcal{W}_w(t)$ for each of the four communities $r\in\{1,2,3,4\}$ and with bin sizes $w\in\{1,3,5,7,9\}$. Panels (a), (b) and (c) indicate results for different thresholds, $\tilde{L}\in\{w,0.8w,0.5w\}$. 

\drt{Our first observation for Fig.~\ref{fig:Time_pic2} is that none of the communities can be detected (for any threshold) if the layers are analyzed in isolation (see results for window size $w=1$). This result is similar to that shown in Fig.~\ref{fig:Time_pic}(b) for summation without thresholding (i.e., whenever $w=1$, we find $|\langle {\bf \hat{v}}^{(r)},{\bf u}^{(r)} \rangle |^2=|\langle {\bf v}^{(r)},{\bf u}^{(r)} \rangle |^2=0$). In other words, the detectability of  communities is only made possible through layer aggregation.}

\drt{Our next observation is that the values $|\langle {\bf \hat{v}}^{(r)},{\bf u}^{(r)} \rangle |^2$ are either zero or close to one, which is in sharp contrast to the values of $|\langle {\bf v}^{(r)},{\bf u}^{(r)} \rangle |^2$ shown in Fig.~\ref{fig:Time_pic}(b), which can be observed to obtain many values across the range $[0,0.7]$. That is, in this experiment the use of thresholding as a nonlinear data filter allows small communities to be either strongly detected or not detected---there is no middle ground for weak detection (which is the case for layer aggregation without thresholding). This is important since error rates for
 community detection vanish as $|\langle {\bf \hat{v}}^{(r)},{\bf u}^{(r)} \rangle |^2\to1$ \cite{Nadakuditi2012}.
}



\drt{Our final observation is that different threshold values enhance the detectability of different communities. For example, 
 community 1 is detectable when $w=3$ for $\tilde{L}\ge 0.8w$ but not for $\tilde{L}=0.5w$ [compare panels (a) and (b) to panel (c)]. Similarly,  community 3 is detectable when $w=9$ for $\tilde{L}\le 0.8w$ but not for $\tilde{L}=w$ [compare panels (b) and (c) to panel (a)].
Interestingly, in this experiment we were able to identify a combination of parameters $(\tilde{L},w)$ that allows accurate detection of all four communities---that is, $|\langle {\bf \hat{v}}^{(r)},{\bf u}^{(r)} \rangle |^2\approx 1$ for bin $\mathcal{W}_w(t)$ only when community $r$ is present in time layer $t$ [i.e., $t\in\mathcal{T}_r$]; otherwise, $|\langle {\bf \hat{v}}^{(r)},{\bf u}^{(r)} \rangle |^2\approx 0$.  We highlight these values of $(\tilde{L},w)$ in panel (b) with a violet box. 
However, we stress that these ``best'' values for $(\tilde{L},w)$ arise in this experiment because the communities are relatively similar in size (i.e., $K_r\in[3,9]$) and density (i.e., $\rho_r=1$).  
In general,  one should not expect there to exist one choice of parameters  $(\tilde{L},w)$ to work well for all communities since the detectability-limit criterion given by Eq.~\eqref{eq:Khat} depends on a complex interplay between the network and community parameters $\{p_l\}$, $\rho_L$, $T_r$, $K_r$, $L$, and $\tilde{L}$.}

\section{Discussion}\label{sec:discussion}

\drt{
There is considerable need to better understand how network preprocessing affects  network-analysis methodologies. 
Herein, we studied how different methods for layer aggregation affect the detectability of small-scale communities in multilayer  networks (including multilayer representations of temporal networks). 
Small-community detection is widely  used for anomaly detection in network data \cite{Alon1998,Nadakuditi2012,Miller2015,Mavroeidis,Ding2012,Chen2013}; in cybersecurity, for example, it allows detection of harmful  events such as attacks  \cite{Mavroeidis}, intrusions \cite{Ding2012}, and fraud \cite{Chen2013}.
Understanding limitations on small-community detection provides insight towards the detectability of these harmful activities. Despite most networks inherently changing in time, previous theory for limitations on small-community detection have been restricted to single-layer networks \cite{Alon1998,Nadakuditi2012} or summation-based aggregation \cite{Nayar2015}.}
We highlight that our  \drt{model and analysis generalizes these} previous works in several ways: 
(i)~\drt{a} community has edge probability $\rho\in(0,1]$ and is not necessarily a clique; 
(ii)~\drt{a} community can persist across a subset of layers; 
(iii) the mean edge probability $p_l$ can \drt{vary across network layers}; and
(iv) the \drt{multilayer/temporal network} can simultaneously contain several communities.

\drt{Thus motivated,} we developed random matrix theory \cite{Benaych_2011,Nadakuditi2012} to analyze \drt{detectability} phase transitions in \drt{which the dominant} eigenvectors of modularity matrices associated with layer-aggregated multilayer networks \drt{localize onto communities, thereby allowing their detection.} We developed theory for when a community with $K_r\ll N$ nodes is hidden (i.e., planted) in $T_r\le L$ layers of a multilayer network with $N$ nodes and $L$ layers.
%
We found a detectability phase transition to occur \drt{for a given community $r$ when its} size $K_r$ surpasses a detectability limit.
\drt{When layers are aggregated by summation, the detectability limit $K_r^*$ is given by  Eq.~\eqref{eq:K} and has the scaling behavior $K_r^*\varpropto \sqrt{NL}/T_r$.
Surprisingly, if $L$ is allowed to vary  this  implies summation-based aggregation enhances community detection even if  the community exists in a vanishing fraction $T_r/L$ of layers, provided that $T_r/L$ decays more slowly than $\mathcal{O}(L^{-1/2})$. This result is surprising since layer aggregation still benefits community detection despite the fact that most layers carry no information about the community.

We also introduced and studied the utility of layer-aggregation with thresholding as a nonlinear data filter to enhance small-community detection. 
%
Our analysis [particularly, Eq.~\eqref{eq:Khat}] revealed that in addition to implementing sparsification and dichotomization, thresholding can allow 
{super-resolution community detection}, whereby the detectability limit decreases by several orders of magnitude  (see Fig.~\ref{fig:Thresholded3}). 
  In particular, we showed in Sec.~\ref{sec:thresh} that $\hat{K}_r^*$ decays exponentially with $\sqrt{L}/T_r$ for clique detection in layer-aggregated sparse networks filtered by threshold $\tilde{L}=T_r$.}


\drt{
To illustrate  practical implications of our results,  in Sec.~\ref{sec:time} we presented an experiment involving the detection of small-communities in a time-varying network, highlighting the following key insights:
\begin{itemize}
\item Aggregating time layers into appropriate-sized bins can allow the detection of small communities that would otherwise be undectable (that is, if the layers were considered in isolation or if all layers were aggregated).
\item Layer aggregation by summation enhances community detection if the community persists across sufficiently many  [specifically, $\mathcal{O}(L^{1/2})$] layers, otherwise it can obscure detection.
\item Layer aggregation with thresholding is a nonlinear data filter that can allow super-resolution community detection of small communities that are otherwise to small for detection.
\item The threshold that best enhances the detection of a small community depends on many parameters, and the detection of multiple communities should, in general, utilize multiple thresholds.
\end{itemize}
%

We have thus provided a theoretical framework supporting how small-community detection in temporal network data can be improved through network preprocessing in which network layers are binned into time windows and are aggregated using summation with thresholding.  This filtering, however, }
should not be approached as a ``one-size-fits-all'' procedure. In particular,  we find there exist optimal time window sizes $w$ and layer-aggregation strategies \drt{that,} in general, are unique to each community (i.e.,  depending on its size, density, persistence across the layers, and etc). While it is important to consider a range of window sizes and layer-aggregation methods, this leads to an unavoidable tradeoff between computational cost and \drt{sufficient} exploration of different parameters.

\drt{
Before concluding, we discuss implications of our work regarding the topic of eigenvector localization in complex networks, 
which is  an important topic in network science \cite{Mendez,Pastor}  for the study of centrality \cite{Martin,Kawamoto,Gleich}, spatial analysis \cite{Cucuringu2}, and core-periphery structure \cite{Barucca,Suweis}. In particular, there is growing interest in extending these ideas to time-varying \cite{Taylor2015} and multilayer networks \cite{Mendez2}.
%
Recently, Ref.~\cite{Murphy2017} showed that an Anderson-localization-type  transition occurs for material transport on several real-world networks (e.g., interconnected ponds of melting sea ice, porous human bone and resistor networks) and noted that they
did not observe the wave interference and scattering effects that typically occur for Anderson localization 
(a widely studied phenomenon in which eigenfunctions localize onto defects in disordered materials \cite{Anderson1961,Abrahams}).
Ref.~\cite{Murphy2017} found the phase transition to coincide with a phase transition in network connectivity 
due to eigenvector localization onto different connected components. Our work complements these findings, showing that a similar localization phenomenon can be onset by small communities---that is, localization does not necessarily require network fragmentation. (We note in passing that connected components can be interpreted as one, and perhaps the strictest, notion of a community.) Future research should further explore the connection between community-based and connected-component-based eigenvector localization on networks, and their relationship to Anderson localization in materials. (See \cite{shi1,shi2} for related research using network-based models for disordered and composite materials.) 
}





 \drt{Finally, we highlight other} extensions to our work that would be interesting to pursue. Motivated by applications for data fusion, recent research \cite{Nayar2015} considered weighted averaging of adjacency matrices, allowing them to optimize the weights for the different network layers. It would be interesting to extend our research to weighted averages, which should be fairly straightforward by redefining $\langle \cdot \rangle$ in Eqs.~\eqref{eq:theta}--\eqref{eq:K} with weights. We leave open the joint optimization of weighting and thresholding. 
%
Finally, it would also be interesting to use our method to study the temporal behavior of communities \cite{Sekara}, such as a set of nodes that form a recurring community in different time windows (i.e., periodically or stochastically).

\acknowledgements{
DT and PJM were supported by the Eunice Kennedy Shriver National Institute of Child Health \& Human Development of the National Institutes of Health (R01HD075712) and a James S. McDonnell Foundation 21st Century Science Initiative - Complex Systems Scholar Award (\#220020315).
RCS was supported by the Assistant Secretary of Defense for Research and Engineering under Air Force Contract No. FA8721-05-C-0002 and/or FA8702-15-D-0001. 
Interpretations, opinions, and conclusions of this work are those of the authors and do not reflect the official
position of these funding agencies.
}

\bibliographystyle{plain}

\begin{thebibliography}{99}

\bibitem{Newman2003} M. E. J. Newman, \emph{The Structure and Function of Complex Networks}, {SIAM Rev.} {\bf45}(2), 167--256 (2003).

\bibitem{Lewis} K. Lewis, J. Kaufman, M. Gonzalez, A. Wimmer and N. Christakis, \emph{Tastes, Ties, and Time: A New Social Network Dataset Using Facebook.com}, Social Networks {\bf30}(4), 330--342 (2008).

\bibitem{Holme2012} P. Holme and J. Saram\"aki, \emph{Temporal Networks}, {Phys. Reports} {\bf519}(3), 97--125 (2012).

\bibitem{Boccaletti2014} S. Boccaletti, G. Bianconi, R. Criado, C. Del Genio, J. G\'omez-Gardenes, M. Romance, I. Sendina-Nadal, Z. Wang and M. Zanin, \emph{The Structure and Dynamics of Multilayer Networks}, {Phys. Reports}, {\bf544}(1), 1--122 (2014).

\bibitem{Kivela2014} M. Kivel\"a, A. Arenas, M. Barthelemy, J. P. Gleeson, Y. Moreno and M. A. Porter, \emph{Multilayer Networks}, {J. of Complex Networks} {\bf2}(3), 203--271 (2014).

\bibitem{Menichetti2014} G. Menichetti, D. Remondini and G. Bianconi, \emph{Correlations Between Weights and Overlap in Ensembles of Weighted Multiplex Networks}, {Phys. Rev. E} {\bf90}(6) 062817 (2014). 

\bibitem{Domenico2015} M. De Domenico, V. Nicosia, A. Arenas and V. Latora, \emph{Structural Reducibility of Multilayer Networks},  {Nat. Comms.} {\bf6}, 6864 (2015).

\bibitem{Kleineberg2016} K. K. Kleineberg, M. Boguna M, M. Serrano and F. Papadopoulos, \emph{Hidden geometric correlations in real multiplex networks},  Nat. Phys. {\bf12}, 1076--1081 (2016).

\bibitem{Stanley2015} N. Stanley, S. Shai, D. Taylor, P. J. Mucha, \emph{Clustering Network Layers with the Strata Multilayer Stochastic Block Model}, IEEE Trans. on Network Science and Engineering {\bf 3}, 95--105 (2016).

\bibitem{Taylor2016} D. Taylor, S. Shai, N. Stanley, P. J. Mucha, \emph{Enhanced Detectability of Community Structure in Multilayer Networks through Layer Aggregation}, Phys. Rev. Lett. {\bf116}, 228301 (2016).

\bibitem{Nayar2015} H. Nayar, B. A. Miller, K. Geyer, R. S. Caceres, S. T. Smith and R. R. Nadakuditi, \emph{Improved Hidden Clique Detection by Optimal Linear Fusion of Multiple Adjacency Matrices}, 49th Asilomar Conf. on Signals, Systems and Computers, 1520 (2015).

\bibitem{Mucha2010} P. J. Mucha, T. Richardson, K. Macon, M. A. Porter and J. P. Onnela, \emph{Community Structure in Time-Dependent, Multiscale, and Multiplex Networks}, {Science} {\bf328}(5980), 876--878 (2010).

\bibitem{Bassett2011}  D. S. Bassett, N. F. Wymbs, M. A. Porter, P. J. Mucha, J. M. Carlson and S. T. Grafton, \emph{Dynamic Reconfiguration of Human Brain Networks During Learning}, {Proc. Natl. Acad. of Sci.} {\bf108}, 7641--8646 (2011).

\bibitem{chung} F. Chung and W. Zhao, \emph{A Sharp PageRank Algorithm with Applications to Edge Ranking and Graph Sparsification}, In Proceedings of the 2010 International Workshop on Algorithms and Models for the Web-Graph, 2--14 (2010).

\bibitem{Hill2016} S. M. Hill and \emph{et al.}, \emph{Inferring Causal Molecular Networks: Empirical Assessment through a Community-Based Effort}, Nat. Methods 13{\bf4}, 310--318 (2016).

\bibitem{Clauset2008} A. Clauset, C. Moore, M. E. J. Newman, \emph{Hierarchical Structure and the Prediction of Missing Links in Network}, Nature {\bf453}(7191), 98--101 (2008).

\bibitem{Newman2017} M. E. J. Newman, \emph{Measurement errors in network data,} Preprint available online at: https://arxiv.org/abs/1703.07376 (2017).


\bibitem{Fortunato} S. Fortunato, \emph{Community Detection in Graphs}, Phys. Reports {\bf486}(3), 75--17 (2010).

\bibitem{Rosvall2008} M. Rosvall and C. T. Bergstrom, \emph{Maps of Random Walks on Complex Networks Reveal Community Structure}, {Proc. Natl. Acad. Sci.} {\bf105}(4),1118 (2008).

\bibitem{Lancichinetti2009} A Lancichinetti, S. Fortunato and F. Radicchi, \emph{Benchmark Graphs for Testing Community Detection Algorithms}, {Phys. Rev. E} {\bf78}(4), 046110 (2008).

\bibitem{Sales2007} M. Sales-Pardo, R. Guimera, A. A. Moreira and L. A. N. Amaral,  \emph{Extracting the hierarchical organization of complex systems}, {Proc. Natl. Acad. Sci.} {\bf104}(39), 15224 (2007).

\bibitem{moody} J. Moody, \emph{Peer Influence Groups: Identifying Dense Clusters in Large Networks}, Social Networks {\bf23}(4), 261--283 (2001).

\bibitem{Mavroeidis} D. Mavroeidis, L. Batina, T. van Laarhoven and E. Marchiori, \emph{PCA, Eigenvector Localization and Clustering for Side-Channel Attacks on Cryptographic Hardware Devices}, In Joint Euro. Conf. on Machine Learning and Knowledge Discovery in Databases, 253--268 (2012).

\bibitem{Ding2012} Q. Ding, N. Katenka, P. Barford, E. Kolaczyk and M. Crovella, \emph{Intrusion as (Anti) Social Communication: Characterization and Detection}, In Proceedings of the 18th ACM SIGKDD International Conference on Knowledge Discovery and Data Mining, 886-894 (2012).

\bibitem{Chen2013} S. Chen and A. Gangopadhyay, \emph{A Novel Approach to Uncover Health Care Frauds Through Spectral Analysis}, In IEEE International Conference on Healthcare Informatics (ICHI), 499--504 (2013).

\bibitem{Alon1998} N. Alon, M. Krivelevich and Benny Sudakov, \emph{Finding a Large Hidden Clique in a Random Graph}, Random Structures and Algorithms {\bf13}, 457--466 (1998).

\bibitem{Nadakuditi2012} R. R. Nadakuditi, \emph{On Hard Limits of Eigen-Analysis Based Planted Clique Detection}, IEEE Statistical Signal Processing Workshop (SSP), 129 (2012).  

\bibitem{Miller2015} B. A. Miller, M. S. Beard, P. J. Wolfe and N. T. Bliss, \emph{Spectral Framework for Anomalous Subgraph Detection}, IEEE Trans. on Signal Processing {\bf63}(16), 4191--4206 (2015).

\bibitem{Ghasemian2015} A. Ghasemian, P. Zhang, A. Clauset, C. Moore and L. Peel, \emph{Detectability Thresholds and Optimal Algorithms for Community Structure in Dynamic Networks}, Phys. Rev. X {\bf6}, 031005 (2016).

\bibitem{Kawamoto_2015} T. Kawamoto and Y. Kabashima, \emph{Detectability of the Spectral Method for Sparse Graph Partitioning}, {EPL (Europhysics Letters) {\bf112}(4), 40007} (2015).


\bibitem{Decelle_2011} A. Decelle, F. Krzakala, C. Moore and L. Zdeborov\'a, \emph{Inference and Phase Transitions in the Detection of Modules in Sparse Networks}, {Phys. Rev. Lett.} {\bf107}(6), 065701 (2011).

\bibitem{Nadakuditi_2012} R. R. Nadakuditi and M. E. J. Newman, \emph{Graph Spectra and the Detectability of cCmmunity Structure in Networks}, {Phys. Rev. Lett.} {\bf108}(18), 188701 (2012).
%
\bibitem{Radicchi_2013} F. Radicchi, \emph{Detectability of Communities in Heterogeneous Networks}, {Phys. Rev. E} {\bf88}(1), 010801 (2013).

\bibitem{Peixoto2013} T. P. Peixoto, \emph{Eigenvalue Spectra of Modular Networks}, {Phys. Rev. Lett.} {\bf111}(9), 098701 (2013).

\bibitem{Chen2015} P. Y. Chen and A. O. Hero, \emph{Phase transitions in Spectral Community Detection}, IEEE Trans. on Signal Processing {\bf63}, 4339--4347 (2015).





\bibitem{Fortunato2007} S. Fortunato and M. Barthelemy, \emph{Resolution Limit in Community Detection}, Proc. of the Natl. Acad. of Sci. {\bf104} 36 (2007).













%
%







\bibitem{Newman_2004} M. E. J. Newman and M. Girvan, \emph{Finding and Evaluating Community Structure in Networks}, {Phys. Rev. E} {\bf69}(2), 026113 (2004).


\bibitem{Bai2010} Z. Bai and J.W. Silverstein, \emph{Spectral Analysis of Large Dimensional Random Matrices} (New York: Springer, 2010).

\bibitem{Benaych_2011} F. Benaych-Georges and R. R. Nadakuditi, \emph{The Eigenvalues and Eigenvectors of Finite, Low Rank Perturbations of Large Random Matrices}, {Adv. in Math.} {\bf227}, 494 (2011).


\bibitem{Capitaine2009} M. Capitaine, C. Donati-Martin and D. F\'eral., \emph{The Largest Eigenvalues of Finite Rank Deformation of Large Wigner Matrices: Convergence and Nonuniversality of the Fluctuations}, Annals of Prob. {\bf37}, 1 (2009). 





\bibitem{Hoeffding} W. Hoeffding, \emph{Probability Inequalities for Sums of Bounded Random Variables}, J. of the Amer. Stat. Assoc. {\bf58}(301), 13--30 (1963).


\bibitem{Mendez} J. A. M\'endez-Berm\'udez, A. Alcazar-Lopez, A. J. Martinez-Mendoza, F. A. Rodrigues and T. K. DM. Peron, \emph{Universality in the Spectral and Eigenfunction Properties of Random Networks}, Phys. Rev. E {\bf91}(3), 032122 (2015).

\bibitem{Pastor} R. Pastor-Satorras and C. Castellano, \emph{Distinct Types of Eigenvector Localization in Networks}, Scientific Reports {\bf6} (2016).

\bibitem{Martin} T. Martin, X. Zhang and M. E. J. Newman, \emph{Localization and Centrality in Networks}, Phys. Rev. E {\bf90}, 052808 (2014).

\bibitem{Kawamoto} T. Kawamoto, \emph{Localized Eigenvectors of the Non-Backtracking Matrix}, J. of Stat. Mech. {\bf2}, 023404 (2016).

\bibitem{Gleich} H. Nassar, K. Kloster and D. F. Gleich, \emph{Strong Localization in Personalized PageRank Vectors}, Algorithms and Models for the Web Graph. Lecture Notes in Computer Science, vol 9479. Springer, In Proceedings of the 2015 Workshop on Algorithms for the Web-Graph, 190--202 (2015).


\bibitem{Cucuringu2} M. Cucuringu, V. D. Blondel and P. Van Dooren, \emph{Extracting Spatial Information from Networks with Low-Order Eigenvectors}, Phys. REv. E {\bf87}(3), 032803 (2013).

\bibitem{Barucca} P. Barucca, D. Tantari and F. Lillo, \emph{Centrality Metrics and Localization in Core-Periphery Networks}, J. of Stat. Mech. {\bf2}, 023401 (2016).

\bibitem{Suweis} S. Suweis, \emph{Effect of Localization on the Stability of Mutualistic Ecological Networks}, Nat. Comms. {\bf6}, 10179 (2015).

\bibitem{Taylor2015} D. Taylor, S. A. Myers, A. Clauset, M. A. Porter and P. J. Mucha, \emph{Eigenvector-Based Centrality Measures for Temporal Networks}, Multiscale Modeling  \& Sim. {\bf15}(1), 537-574 (2017).

\bibitem{Mendez2} J. A. M\'endez-Berm\'udez, G. F. de Arruda, F. A. Rodrigues and Y. Moreno, \emph{Scaling Properties of Multilayer Random Networks},  arXiv preprint arXiv:1611.06695 (2016).


\bibitem{Murphy2017} N. B. Murphy, E. Cherkaev and K. M. Golden, \emph{Anderson Transition for Classical Transport in Composite Materials}, Phys. Rev. Lett. {\bf118}, 036401 (2017). 


\bibitem{Anderson1961} P. W. Anderson, \emph{Localized Magnetic States in Metals}, Phys. Rev. Lett., {\bf124}, 41 (1961).

\bibitem{Abrahams} E. Abrahams, P. W. Anderson, D.C. Licciardello and T. V. Ramakrishnan, \emph{Scaling Theory of Localization: Absence of Quantum Diffusion in Two Dimensions}, Phys. Rev. Lett. {\bf42}(10) 673 (1979).




\bibitem{shi1} F. Shi, S. Wang, M. G. Forest and P. J.  Mucha, \emph{Percolation-Induced Exponential Scaling in the Large Current Tails of Random Resistor Networks,} Multiscale Modeling \& Sim., {\bf11}(4), 1298-1310 (2013).

\bibitem{shi2} F. Shi, S. Wang, M. G.  Forest, P. J. Mucha and R. Zhou, \emph{Network-Based Assessments of Percolation-Induced Current Distributions in Sheared Rod Macromolecular Dispersions}, Multiscale Modeling \& Sim., {\bf12}(1), 249-264 (2014).

\bibitem{Sekara}  V. Sekara, A. Stopczynski and S. Lehmann, \emph{Fundamental Structures of Dynamic Social Networks}, Proc. Natl. Acad. of Sci.  {\bf113}(36), 9977--9982 (2016).


\end{thebibliography}

\end{document}